\documentclass[aps,pra,superscriptaddress,nofootinbib,twocolumn]{revtex4-1}

\usepackage[T1]{fontenc}
\usepackage[utf8]{inputenc}

\usepackage{float}
\usepackage{graphicx}
\usepackage{epsfig,epstopdf}
\usepackage[table]{xcolor}
\usepackage{xcolor}
\usepackage{tikz}
\usepackage{pgffor}
\usepackage{soul}

\usepackage[english]{babel}

\PassOptionsToPackage{hyphens}{url}
\usepackage[hyphenbreaks]{breakurl}

\usepackage{times}
\usepackage{bbm}
\usepackage{amsthm}
\usepackage{mathbbol}
\usepackage{mathtools}
\usepackage{SIunits}

\usepackage{enumerate}
\usepackage{enumitem}
\usepackage{booktabs}
\usepackage{xprintlen}

\usepackage{bibunits}
\defaultbibliography{supremacy_dom.bib,ion_trap_speedups.bib}
\defaultbibliographystyle{./myapsrev4-1}

\usepackage{pgfplots}
\usetikzlibrary{pgfplots.groupplots}
\pgfplotsset{compat=1.11}
\usepgfplotslibrary{fillbetween}

\usepackage{tikz}
\usetikzlibrary{
  shapes,
  shapes.geometric,
	trees,
	matrix,
  positioning,
through,
calc,
    pgfplots.groupplots,
  }

\PassOptionsToPackage{pdftex,hyperfootnotes=false,pdfpagelabels}{hyperref}
\usepackage{hyperref}
\hypersetup{%
    colorlinks=true, linktocpage=true, pdfstartpage=3, pdfstartview=FitV,%
    breaklinks=true, pdfpagemode=UseNone, pageanchor=true, pdfpagemode=UseOutlines,%
    plainpages=false, bookmarksnumbered, bookmarksopen=true, bookmarksopenlevel=1,%
    hypertexnames=true, pdfhighlight=/O,
    urlcolor=blue!70!black, linkcolor=blue!70!black, citecolor=black!50, 
}   
\usepackage[capitalize,compress]{cleveref}

\makeatletter 
\hypersetup{
    pdftitle = {Verifiable measurement-based quantum random sampling with trapped ions},
    pdfauthor = {
      Martin Ringbauer, 
      Marcel Hinsche, 
      Thomas Feldker, 
      Paul K. Faehrmann,  
      Juani Bermejo-Vega, 
      Claire Edmunds, 
      Roman Stricker, 
      Christian D. Marciniak, 
      Michael Meth, 
      Ivan Pogorelov, 
      Lukas Postler, 
      Rainer Blatt, 
      Philipp Schindler, 
      Jens Eisert, 
      Thomas Monz, 
      Dominik Hangleiter
},
    pdfsubject = {Quantum computation, quantum information},
    pdfkeywords = {Quantum advantage,
        quantum supremacy,
        random circuit sampling, 
        quantum random sampling, 
        trapped ions, 
        benchmarking,
        verification,
        quantum processors,
        qubit recycling,
         }
    }
\makeatother

\DeclareMathOperator{\tr}{Tr}
\newcommand{\ket}[1]{\vert{#1}\rangle}
\newcommand{\bra}[1]{\langle{#1}\vert}
\newcommand{\braket}[2]{\langle{#1}\vert #2 \rangle}
\newcommand{\proj}[1]{\ket{#1}\bra{#1}}

\newcommand{\ms}{\text{MS}}

\DeclareMathOperator{\var}{Var}

\DeclareMathOperator{\diag}{diag}

\DeclareMathOperator{\poly}{poly}
\DeclareMathOperator{\Eb}{\mathbb{E}}

\newcommand{\mc}[1]{\mathcal #1}
\newcommand{\mb}[1]{\mathbb #1}

\newcommand{\lint}{{\textsf{lin}}}
\newcommand{\logt}{{\textsf{log}}}

\newcommand{\id}{\mathbbm{1}}
\newcommand{\ee}{\mathrm{e}}
\newcommand{\ii}{\mathrm{i}}

\usepackage{xcolor}

\definecolor{christian}{rgb}{0,.4,1}

\definecolor{jens}{rgb}{0,.8,.5}

\definecolor{juan}{rgb}{.7,.1,0}

\definecolor{dominik}{rgb}{0.4,.0,0.6}

\definecolor{dominik}{rgb}{0.4,.0,0.6}

\definecolor{paul}{rgb}{0.4,.5,0.6}

\definecolor{tm}{rgb}{0,1,0}

\definecolor{mr}{rgb}{0,.7,0}

\newcommand{\innsbruck}{Universit\"{a}t Innsbruck, Institut f\"{u}r Experimentalphysik, Technikerstrasse 25, 6020 Innsbruck, Austria}

\newcommand{\aqt}{Alpine Quantum Technologies GmbH, 6020 Innsbruck, Austria}

\newcommand{\iqoqi}{Institut f\"{u}r Quantenoptik und Quanteninformation, \"{O}sterreichische Akademie der  Wissenschaften, Otto-Hittmair-Platz 1, 6020 Innsbruck, Austria}

\newcommand{\fu}{Dahlem Center for Complex Quantum Systems, Freie Universit{\"a}t Berlin, 14195 Berlin, Germany}

\newcommand{\quics}{Joint Center for Quantum Information and Computer Science (QuICS), University of Maryland \& NIST, College Park, MD 20742, USA}

\newcommand{\jqi}{Joint Quantum Institute (JQI), University of Maryland \& NIST, College Park, MD 20742, USA}
\newcommand{\hzb}{Helmholtz-Zentrum Berlin f{\"u}r Materialien und Energie, 14109 Berlin, Germany}

\newcommand{\hhi}{Fraunhofer Heinrich Hertz Institute, 10587 Berlin, 
Germany}

\begin{document}

\begin{bibunit}
\title{Verifiable measurement-based quantum random sampling with trapped ions}

\author{Martin Ringbauer}
\email{martin.ringbauer@uibk.ac.at}
\affiliation{\innsbruck}
\author{Marcel Hinsche}
\affiliation{\fu}
\author{Thomas Feldker}
\affiliation{\innsbruck}
\affiliation{\aqt}
\author{Paul K. Faehrmann}
\affiliation{\fu}
\author{Juani Bermejo-Vega}
\affiliation{\fu}
\affiliation{Departamento  de Electromagnetismo y Física de la Materia, Avenida de la Fuente Nueva, 18071 Granada, Universidad de Granada,  Granada, Spain}
\affiliation{Institute Carlos I for Theoretical and Computational Physics, Campus Universitario Fuentenueva,
Calle Dr. Severo Ochoa, 18071,
Granada, Spain.}
\author{Claire Edmunds}
\affiliation{\innsbruck}
\author{Lukas Postler}
\affiliation{\innsbruck}
\author{Roman Stricker}
\affiliation{\innsbruck}
\author{Christian D. Marciniak}
\affiliation{\innsbruck}
\author{Michael Meth}
\affiliation{\innsbruck}
\author{Ivan Pogorelov}
\affiliation{\innsbruck}
\author{Rainer Blatt}
\affiliation{\innsbruck}
\affiliation{\aqt}
\affiliation{\iqoqi}
\author{Philipp Schindler}
\affiliation{\innsbruck}
\author{Jens Eisert}
\affiliation{\fu}
\affiliation{\hzb}
\affiliation{\hhi}
\author{Thomas Monz}
\affiliation{\innsbruck}
\affiliation{\aqt}
\author{Dominik Hangleiter}
\email{mail@dhangleiter.eu}
\affiliation{\quics}
\affiliation{\jqi}

\date{\today}

\begin{abstract}
 Quantum computers are now on the brink of outperforming their classical counterparts. One way to demonstrate the advantage of quantum computation is through quantum random sampling performed on quantum computing devices.
However, existing tools for verifying that a quantum device indeed performed the classically intractable sampling task are either impractical or not scalable to the quantum advantage regime.
The verification problem thus remains an outstanding challenge.
Here, we experimentally demonstrate efficiently verifiable quantum random sampling in the measurement-based model of quantum computation
on a trapped-ion quantum processor.
We create and sample from random cluster states, which are at the heart of measurement-based computing, up to a size of $4\times 4$ qubits. By exploiting the structure of these states, we are able to recycle qubits during the computation to sample from entangled cluster states that are larger than the qubit register. We then efficiently estimate the fidelity to verify the prepared states---in single instances and on average---and compare our results to cross-entropy benchmarking.
Finally, we study the effect of experimental noise on the certificates. Our results and techniques provide a feasible path toward a verified demonstration of a quantum advantage.
\end{abstract} 

\maketitle


In \emph{quantum random sampling}, a quantum device is used to produce samples from the probability distribution generated by a random quantum computation \cite{hangleiter_computational_2023}. 
This is a particularly challenging task for a classical computer asymptotically~\cite{aaronson_computational_2013,bremner_average-case_2016,bouland_complexity_2019} and in practice~\cite{pan_solving_2022,kalachev_classical_2021} and thus at the center of recent demonstrations of a quantum advantage~\cite{arute_quantum_2019_short,zhu_quantum_2022_short,zhong_phase-programmable_2021,madsen_quantum_2022}. 
A key challenge for such experiments, however, is to verify that the produced samples indeed originate from the probability distribution generated by the correct random quantum computation. 
Verification based only on classical samples from the device is fundamentally inefficient~\cite{hangleiter_sample_2019}. 
In practice, the verification problem has been approached using so-called linear \emph{cross-entropy benchmarking} (XEB)~\cite{boixo_characterizing_2018,arute_quantum_2019_short}. 
The corresponding \emph{XEB score} is  obtained by averaging the ideal probabilities corresponding to the observed experimental samples. 
XEB is appealing since it has been argued that even achieving any non-trivial XEB score might be a classically computationally intractable task~\cite{aaronson_complexity-theoretic_2017,aaronson_classical_2020} and that it can be used to sample-efficiently estimate the \emph{quantum fidelity} of the experimental quantum state~\cite{arute_quantum_2019_short,choi_emergent_2021}.
However, XEB requires a classical simulation of the implemented circuits to obtain the ideal output distribution. 
The computational run-time of estimating XEB from samples thus scales exponentially, rendering it practically infeasible in the quantum advantage regime. 
Moreover, it is not always a good measure of the quantum fidelity~\cite{gao_limitations_2021,ware_sharp_2023,morvan_phase_2023}.
Another way classical verification of quantum devices has been approached is via interactive proof systems \cite{brakerski_cryptographic_2018,mahadev_classical_2018}, albeit at the cost of large device overheads \cite{zhu_interactive_2021-1,stricker_towards_2022}.
Hence, classical approaches to verification have limited applicability for devices operating in the quantum advantage regime. 

\begin{figure*}[t]\includegraphics[width=\linewidth]{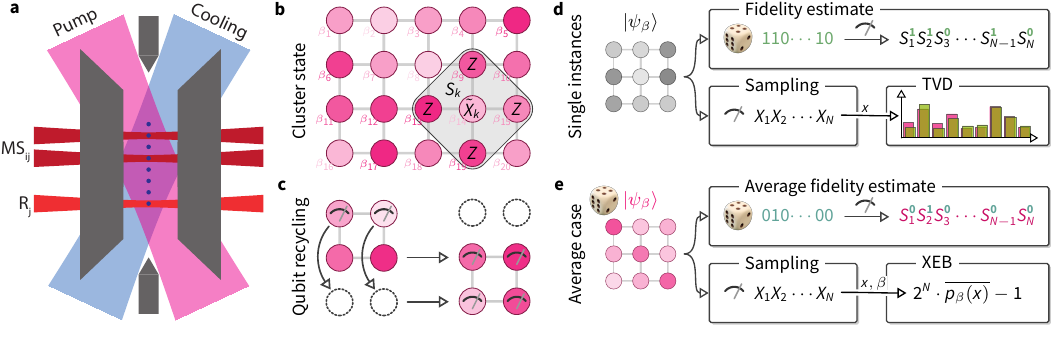}
  \caption{\label{fig:protocol} \textbf{Overview of the experiment.} \textbf{(a) Sketch of the ion trap quantum processor.} Strings of up to 16 ions are trapped in a linear chain. Any single ion or pair of ions can be individually addressed by means of steerable, tightly focused laser beams (dark red) to apply resonant operations $\mathrm{R}_j$ or M\o{}lmer-S\o{}rensen entangling gates $\mathrm{MS}_{i,j}$. Global detection, cooling (blue), and repumping (pink) beams are used to perform a mid-circuit reset of part of the qubit register~\cite{Ringbauer2021}.
  \textbf{(b) Implemented cluster states.} Cluster states with local rotation angles $\beta_{i} \in \{0, \frac \pi 4, \ldots, \frac{7\pi} 4\} $ up to a size of $4 \times 4$ qubits are created in the qubit register.
  Each cluster state is defined by its $N$ stabilizers $S_k$ which are given by rotated $X$ operators $\tilde X_k = X_k(\beta_k)$ at each site $k = 1, \ldots, N$ multiplied with $Z$ operators on the respective neighbouring sites.
  \textbf{(c) Recycling of qubits.}
  Using sub-register reset of qubits, we prepare cluster states that are larger than the qubit register. For example, using four ions, we prepare cluster states of size $2 \times 3$. 
  \textbf{(d) Single-instance verification.}
  In order to verify single cluster state preparations with fixed rotation angles $\beta$, we measure it in different bases.
  To perform fidelity estimation we measure uniformly random elements of its stabilizer group, which is obtained by drawing a random product of the $N$ stabilizers $S_k$, indexed by a length-$N$ random bitstring indicating for each $S_k$ whether it participates in the product.
  To sample from the output distribution, we measure in the $X$-basis. These samples are verified in small instances by the empirical total-variation distance (TVD). 
  \textbf{(e) Average-case verification.}
  To assess the average quality of the cluster state preparations, we perform measurements on cluster states with random rotations. 
  By measuring a random element of the stabilizer group of each random cluster state, we obtain an estimate of the average fidelity. 
  From the samples from random cluster states in the $X$-basis, we compute the \emph{cross-entropy benchmark} (XEB) by averaging the ideal probabilities $p_\beta(x)$ corresponding to the samples $x$ and the cluster with angles $\beta$. 
  }
\end{figure*}

These challenges raise the question of whether there are quantum verification techniques that could be used to \emph{efficiently verify} quantum random sampling experiments, even when their simulation is beyond the computational capabilities of classical devices. 
Answering this question in the affirmative, we turn to a different universal model of quantum computation---\emph{measurement-based quantum computing} (MBQC)~\cite{raussendorf_one-way_2001,raussendorf_measurement-based_2003}. 
In contrast to the circuit model, a computation in MBQC proceeds through measurements, instead of unitary operations, applied sequentially to an entangled \emph{cluster state}~\cite{raussendorf_measurement-based_2003}. 
Roughly speaking, a cluster state on an $n \times m$ grid of qubits can be used to execute an $n$-qubit, depth-$m$ quantum circuit.
 %
Appropriately randomized, cluster states turn out to be a source of random samples appropriate for demonstrating a quantum advantage via random sampling~\cite{gao_quantum_2017,bermejo-vega_architectures_2018,haferkamp_closing_2020}. 
Crucially, though, each cluster state is fully determined by a small set of so-called \emph{stabilizer operators}. 
By measuring the stabilizer operators using well-characterized single-qubit measurements, preparations of these cluster states can be efficiently verified~\cite{flammia_direct_2011,hangleiter_direct_2017,Saggio2018,dangniam_optimal_2020,hangleiter_sampling_2021,tiurev_fidelity_2021}.

Here, we experimentally demonstrate efficiently verifiable quantum random sampling in the MBQC model in two \emph{trapped-ion quantum processors} (TIQP). 
While cluster state generation in TIQP has previously been limited to a size of $2\times 2$~\cite{lanyon_measurement-based_2013}, we overcome this limitation with a two-fold approach. 
First, we use pairwise addressed M\o{}lmer-S\o{}rensen entangling operations~\cite{Ringbauer2021,Pogorelov2021} in a fully connected linear chain to enable the efficient generation of clusters up to a size of $4\times 4$ qubits. 
Second, we make use of spectroscopic decoupling and optical pumping~\cite{schindler_quantum_2013} to perform mid-circuit readout and reset of qubits in order to recycle them. 
In this way, we are able to sequentially measure rows of the cluster and then reuse the measured qubits to prepare a new row of the cluster, while maintaining entanglement with the remaining qubits, see \cref{fig:protocol}(c). 
This allows us to sample from a cluster state on a lattice that is larger than the size of the physical qubit register.
This combination of techniques provides a feasible path towards generating large-scale entangled cluster states using trapped ions.

We then estimate the fidelity of the experimental cluster states in order to verify those states. 
Specifically, we apply a novel variant of \emph{direct fidelity estimation}~\cite{flammia_direct_2011,hangleiter_sampling_2021} to estimate the single-instance fidelity of a fixed cluster state, and the average fidelity of random cluster states. 
The single-instance fidelity certifies the samples from a fixed, random cluster state, and therefore a quantum advantage for sufficiently large cluster states~\cite{hangleiter_direct_2017}. 
Conversely, the average fidelity of random cluster states is a benchmark of the average performance of the quantum processor in the quantum advantage regime~\cite{eisert_quantum_2020}. 
Direct (average) fidelity estimation is therefore a unified framework for verification and benchmarking of MBQC, analogously to XEB.
However, in contrast to XEB, the fidelity estimation approach has several major advantages:  
First, it is efficient in terms of both the required number of experiments, and the complexity of the postprocessing. 
Second, it requires knowledge only of the measurement noise as opposed to the noise properties of all gates which is required for XEB \cite{gao_limitations_2021,ware_sharp_2023,morvan_phase_2023}.
Finally, the fidelity gives a rigorous bound on the quality of the samples from a single quantum state, whereas XEB is generally only accurate on average. 

In order to assess the performance of the fidelity-derived certificates, we compare them to the available---but inefficient---classical means of certification of the samples,   which is still possible in our proof-of-principle demonstration. In the single-instance case, we compare the experimental performance of the single-instance fidelity estimate to the empirical total-variation distance of the sampled distribution.
In the average case, we compare the average fidelity estimate to the average XEB score. 
Additionally, we study the effect of native noise sources on the different measures of quality. 

Our work thus provides a clear and feasible path towards verified quantum advantage. 
It does so by developing a new approach to verifying random cluster states based on a variant of direct fidelity estimation, introducing the use of qubit recycling in order to generate large clusters, and demonstrating the feasibility of the proposed techniques in the presence of experimental noise.

\begin{figure*}
  \includegraphics[width=\linewidth]{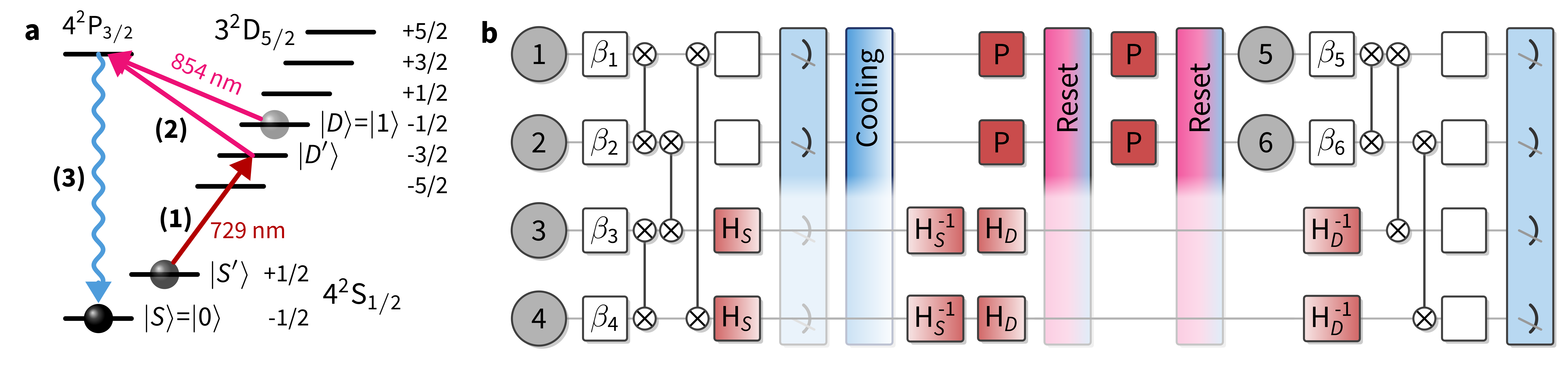}
  \caption{\label{fig:reset} \textbf{Sketch of the circuit with qubit recycling.} \textbf{(a) Recycling.} After detection, a measured qubit is either still in the $\ket{1}$ state (``dark'' outcome) or in one of the two S levels (``bright'' outcome). We reset it to the $\ket{0}$ state by first applying an addressed $\pi$-pulse (1) on the $\ket{S'}\rightarrow\ket{D'}$-transition. A subsequent global \unit{854}{\nano\meter} quench pulse (2) transfers population from all D-levels to the P$_{3/2}$ manifold, from which (3) spontaneous decay occurs, preferentially to the $\ket{0}$ state in the S manifold. We repeat this process twice, which is sufficient to return about 99\% of the population to the $\ket{0}$ state.
  \textbf{(b) Circuit.} The individual qubits are prepared in a product state depending on the random angles $\beta_i$ and entangled via $XX$ interactions and some single-qubit gates (white boxes) to create a cluster state; see the SI, \cref{app:compiled circuits} for details. 
  The measurement of the qubits is achieved by exciting the $P \leftrightarrow S$ transition. 
  In order to perform a circuit with recycling, a coherent $\pi$-pulse on the $S \rightarrow D'$ transition (denoted by H$_S$) is applied to `hide' the qubits which should not be measured in the D-manifold. After the measurement, the chain is cooled using polarization-gradient cooling. The reset makes use of local pulses on the measured qubit that transfer the remaining population of $\ket {S'}$ to the D$_{5/2}$-manifold (denoted by P) and global pulses that transfer the population of that manifold back to $\ket 0 $. 
  Prior to the reset, all unmeasured qubits are `hidden' in the S$_{1/2}$-manifold. For this, the population which was in $\ket 0$ prior to the measurement is coherently transferred back to $\ket S$ via a $\pi$-pulse (H$_S^{-1}$), and the population which is in $\ket 1$ is transferred to $\ket {S'}$ via a $\pi$-pulse on the $D\rightarrow S'$ transition (H$_D$). After the reset procedure (a), a $\pi$-pulse (H$_D^{-1}$) is applied to the unmeasured qubits to transfer the population which was previously in $\ket 1$ back from~$S'$.}
\end{figure*}

\textbf{Sampling and verification protocols.} 
In the circuit model, natural examples of random computations are, for instance, circuits composed of Haar-random two-qubit gates \cite{brandao_local_2016}, or composed of native entangling gates and random single-qubit gates~\cite{arute_quantum_2019_short}. 
In contrast, in MBQC, a universal quantum computation can be realized by adaptively performing single-qubit rotations around the $Z$-axis on a cluster state and measuring in the Hadamard basis conditioned on the outcomes of previous Hadamard-basis measurements~\cite{raussendorf_measurement-based_2003,mantri_universality_2017,haferkamp_closing_2020}.
This leads to a natural notion of random MBQC wherein those single-qubit $Z$-rotations are applied with angles chosen randomly from an appropriate discretization of the unit circle~\cite{bermejo-vega_architectures_2018,haferkamp_closing_2020}. 
Adaptively performing single-qubit rotations then becomes superfluous since they are chosen randomly anyway, and an outcome pattern on the square lattice defines both, an effective quantum circuit given the random rotations, and the outcomes of measuring that circuit. 
Hence, repeatedly measuring a fixed, random cluster state without adaptive rotations is equivalent to measuring many different quantum circuits chosen randomly from an ensemble defined by the random rotations, see \cref{fig:physical logical circuits} of the Supplementary Information (SI) for details. 

The largest discretization in the choice of single-qubit rotations leading to a computationally universal MBQC scheme consists of eight evenly spaced angles, corresponding to powers of the $T$ gate.
In exactly the same way as for circuit-based sampling schemes \cite{bremner_average-case_2016,boixo_characterizing_2018}, there is strong complexity-theoretic evidence that for $m \gtrsim n$ approximately reproducing the outcome statistics of such random measurements is classically  intractable~\cite{bermejo-vega_architectures_2018,haferkamp_closing_2020}. 
In fact, in both cases, even producing samples from a quantum state with a non-vanishing or only slowly vanishing fidelity is likely classically hard~\cite{gao_limitations_2021,aharonov_polynomial-time_2022}.
Quantum advantage aside, the effective computations implemented by random cluster states generate a unitary $2$-design~\cite{haferkamp_closing_2020} and therefore yield a reliable average-performance benchmark for measurement-based computations~\cite{heinrich_general_2022}.

Concretely, the MBQC random sampling protocol we apply is then the following~\cite{bermejo-vega_architectures_2018,haferkamp_closing_2020} (see \cref{fig:protocol}, and \cref{app:compiled circuits} of the SI for explicit circuits): 
\begin{enumerate}
  \item Prepare a cluster state on $N = n \times m $ qubits on a rectangular lattice by preparing each qubit in the $\ket +$ state and applying controlled-$Z$ gates between all neighbors. 
  \item Apply single-qubit rotations $Z(\beta) = \ee^{- \ii \beta Z/2}$ with random angles $\beta \in \{ 0, \frac \pi 4, \ldots, \frac{7\pi}4\}$ to every qubit. 
  \item Measure all qubits in the Hadamard basis.
\end{enumerate}
We note that the state preparation steps 1 and 2 can also be achieved by time-evolving an initial state $\ket +^{\otimes N}$ under an Ising Hamiltonian on an $n \times m $ lattice with random local fields depending on the $\beta_k$~\cite{bermejo-vega_architectures_2018,haferkamp_closing_2020}, but the gate-based approach outlined here is more suitable for TIQP.

Using a variant of direct fidelity estimation (DFE), we assess the quality of both single cluster states with local $Z$ rotations, and the average quality of such state preparations.
In DFE, we estimate the fidelity $F(\rho, \proj \psi)  = \bra \psi \rho \ket \psi$ of a fixed experimental state $\rho$ by measuring random operators from the \emph{stabilizer group} of the random cluster state $\ket \psi$ and averaging the results, see Methods for detail.
The stabilizer group is the group generated by the $N$ stabilizers of the random cluster. 
Each stabilizer $S_k$ is the product of a rotated $X$-operator at site $k$---given by $X_k(\beta) = \ee^{-\ii \beta X_k/2}$---and $Z$-operators on the neighboring sites, giving rise to a characteristic star shape on the square lattice, see \cref{fig:protocol}(b). 
Importantly, all elements of the stabilizer group are products of single-qubit operators. 
Our trust in the fidelity estimate therefore only depends on our ability to reliably perform single-qubit measurements, which we verify.  
In order to measure the average fidelity over the set of cluster states, we prepare random cluster states and for each state measure a random element of its stabilizer group. 
We then average the results to obtain an estimate of the average fidelity. 
At a high level, fidelity estimation thus exploits our ability to measure the experimental state in different bases. 
It requires a number of experimental state preparations that is \emph{independent} of the size of the system, making it scalable to arbitrary system sizes, see Methods for details.
We note that we also measured a witness for the fidelity~\cite{hangleiter_direct_2017} and find that it is not practical in a scalable way for noisy state preparations, as we detail in \cref{app:witness} of the SI.

Given the relatively small system sizes of the experiments in this work, we are also able to directly compute non-scalable measures of quality that make use of the classical samples only.
This enables us to compare fidelity estimation with inefficient classical verification methods in different scenarios.
To classically assess the quality of samples from a fixed experimental state preparation, we use the \emph{total-variation distance} (TVD) ${d_{\text{TV}}(P,Q) = \sum_{x} | P(x) - Q(x)|/2}$. The TVD quantifies the optimal probability of distinguishing the experimentally sampled distribution $Q$ and the ideal one for a noiseless cluster $P$. 
The TVD is the classical analog of the trace distance $d_{\tr}(\rho, \proj \psi ) = \tr(|\rho - \proj \psi|)/2$, which quantifies the optimal probability of distinguishing the sampled quantum states $\rho$ and $\proj \psi$.
The fidelity $F$ upper-bounds the trace distance~\cite{fuchs_cryptographic_1999} and therefore the TVD of the sampled distributions as
\begin{align}
\label{eq:estimate tvd relation}
  d_{\text{TV}} \leq d_{\tr} \leq \sqrt{1-F} .
\end{align}
The root infidelity $\sqrt{1-F}$ can therefore be used to certify the classical samples from $\rho$. 
We note that it is a priori not clear how tight this bound is in an experimental scenario and how experimental noise affects the different verification methods.  
In order to classically assess the average quality of the quantum device, we estimate the linear XEB fidelity between $Q$ and $P$, which is defined as $f_\lint(Q,P) = 2^n \sum_x Q(x) P(x) - 1 $~\cite{hangleiter_computational_2023}. 
The average XEB fidelity over the random cluster states measures the average quantum fidelity in the regime of low noise \cite{gao_limitations_2021,ware_sharp_2023,morvan_phase_2023}, see \cref{app:average fidelity} of the SI for details.
\smallskip

\textbf{Experimental implementation.} 
We implement the random MBQC sampling and verification protocols in two ion-trap quantum processors. Quantum information is encoded in the S$_{1/2}$ ground state and D$_{5/2}$ excited state of up to 16 $^{40}$Ca$^+$ ions confined in a linear Paul trap~\cite{schindler_quantum_2013,Pogorelov2021}. 
We use these devices to implement two sets of experiments. 
First, we generate rectangular $n \times m$ random cluster states of up to 16 ions by appropriately entangling the respective ions in a linear chain using pairwise addressed M{\o}lmer-S{\o}rensen-gates~\cite{Ringbauer2021,Pogorelov2021}. 
In a second, proof-of-principle set of experiments on a device with an extended control toolbox yet somewhat lower fidelities, we make use of spectroscopic decoupling and optical pumping to recycle qubits to demonstrate a more qubit-efficient way to sample from large-scale entangled cluster states. 
By construction, the 2D cluster states require entangling gates between neighboring qubits only. 
As a consequence, when generating the cluster from top to bottom, once the first row has been entangled to the second, we can measure the qubits of the first row. 
Once measured, these qubits can be reset to the ground state, prepared in their appropriate initial states and entangled as the third row of the cluster state, and so on. 
Due to the local entanglement structure of the cluster state, the measurement statistics obtained in this way are identical to the statistics that would be obtained from preparing and measuring the full cluster state at once.

Experimentally, we make use of mid-circuit readout capabilities~\cite{Stricker2020} using an EMCCD camera to read out a subset of the qubits, while spectroscopically decoupling the remaining qubits from the readout beams, see \cref{fig:reset}. 
After the readout, we re-cool the ion string using a combination of Doppler cooling and polarization-gradient cooling for a total of \unit{3}{\milli\second}. 
Then we employ two rounds of optical pumping using addressed \unit{729}{\nano\meter} pulses in combination with a global \unit{854}{\nano\meter} quench beam to reset the qubits to the $\ket{0}$ ground state~\cite{schindler_quantum_2013}, while the remaining qubits are spectroscopically decoupled. 
This completes the reset and we can now prepare the measured qubits in their new states and entangle them to the remaining qubits of the cluster, see \cref{fig:reset}. This procedure enables us to sample from entangled quantum states with more qubits than the physical register size of the used quantum processor. 
Specifically, to prepare an $n \times m$ cluster state at least $n+1$ repeatedly recycled qubits are required, and the required circuit depth (and recycling steps) decreases as the number of available physical qubits increases. 

For every state, we perform sampling and verification measurements. 
We measure the state in the Hadamard basis in order to perform sampling. 
For verification, we measure a random element of its stabilizer group.
When verifying a single instance of a state preparation, we repeat this procedure for a fixed state and then estimate the fidelity from the random stabilizer measurements and the TVD from the classical samples. 
To estimate the average performance of the device, we repeat the procedure for random states and estimate the average fidelity and the average XEB fidelity, see \cref{fig:protocol}(d,e). 
Finally, for the $2\times 2$ cluster, we study the effect of increasing global (local) dephasing noise on the verification performance by adding small (un)correlated random $Z$-rotations before and after each entangling gate.
  \smallskip

\begin{figure}
  \includegraphics{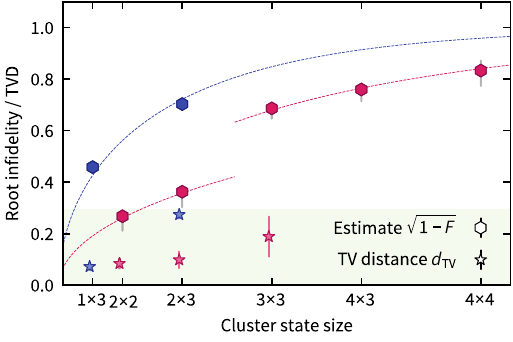}
  \caption{\label{fig:results} 
  \textbf{Experimental results for single-instance verification.}
  Root infidelity estimate $\sqrt{1- F}$ (hexagons), and empirical TVD (stars) for single instances of random MBQC cluster states with recycling (blue) and without (pink). Note that the horizontal axis is labelled with the cluster size $n\times m$ and scaled with qubit number $n\, m$.
  The root infidelity upper-bounds the TVD per \cref{eq:estimate tvd relation}.
  Colored error bars represent the 3$\sigma$ interval of the statistical error.
  Uncorrelated measurement noise reduces or increases the measured state fidelity compared to the true fidelity asymmetrically depending on its value, such that the shown values are lower bounds to the true state fidelity, see the Methods section for details. 
  The worst-case behaviour of the measurement noise is represented by gray error bars.
  In the non-recycling experiment, the register size is increased between the $2\times 3$ and the $3\times 3$ instance, leading to a decrease in the local gate fidelities. 
  Modeling the noise as local depolarizing noise after each entangling gate (dotted lines), we obtain effective local Pauli error probabilities after the two-qubit gates of 5.3\%, 2.6\%, and 1.0\%, for the recycling data, for the large-register non-recycling data, and the small register non-recycling data, respectively; see \cref{app:average fidelity} of the SI. 
  The shaded green area is the acceptance region corresponding to an infidelity threshold of 8.6\% arising in the rigorous hardness argument as sketched in \cref{app:stockmeyer argument} of the SI.
  Since the accuracy of the TVD estimate scales with the system dimension already for cluster sizes of $4\times3$ and $4\times4$ infeasible amount of samples would be required for an accurate estimate, and hence these are not shown.
  See \cref{tab:num samples} of the SI for experimental details. 
  }
\end{figure}

\begin{figure}[t]
  \includegraphics{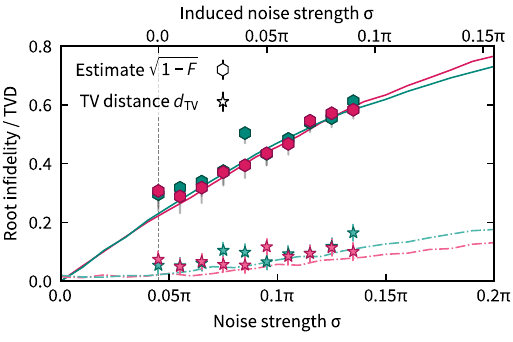}
  \caption{
  \label{fig:noise} 
  \textbf{Single-instance verification with artificially added phase noise.} Root infidelity estimate $\sqrt{1-F}$ (hexagons) and empirical total-variation distance $d_{\text{TV}}$ (stars) of a $2\times 2$ cluster state with artificially introduced local (pink) and global (green) phase noise---$Z$-rotations with rotation angle drawn from a Gaussian distribution with variance $\sigma^2$---before and after M\o{}lmer-S\o{}rensen gate applications as a function of the noise strength $\sigma$, see Methods for details.
  Solid (dashed) lines show simulated root infidelity (total-variation distance) for the respective types of noise. 
  The experimental data (top axis) is shifted with respect to the simulations (bottom axis) due to the fact that there is residual noise when no artificial noise is introduced. The value of the relative shift given by $0.045\pi$ (dashed vertical line) provides an estimate for the natural noise strength. 
  Colored error bars represent the 3$\sigma$ interval of the statistical error.
  The systematic measurement error of the fidelity estimate is represented by gray error bars. 
  }
\end{figure}
\textbf{Results.} 
We first measure the fidelity and TVD of single random cluster state preparations for various cluster sizes. The results demonstrate that the root-infidelity provides meaningful upper bounds on the TVD, see \cref{fig:results}. 
Importantly, while the efficiently measurable and computable root infidelity estimate is guaranteed to bound the TVD per \cref{eq:estimate tvd relation}, these scalable bounds are not tight. This is seen in \cref{fig:results} as a gap between the root infidelity upper bound and the measured TVD values. 
Indeed, it is expected that reproducing the full quantum state (as measured by the fidelity) is a more stringent requirement than merely reproducing the outcome distribution in one particular measurement basis (as measured by the TVD). 
Hence, the efficient quantum methods require higher fidelities for the corresponding certificate to meet the quantum advantage threshold. 
Notably, above the cluster size of $3\times 3$ qubits, empirically estimating the TVD with sufficient accuracy is practically infeasible due to the exponentially growing state space.
In the proof-of-principle experiments, where recycling is used, we see the same qualitative behavior, although the overall root infidelities are higher. 
This is likely due to imperfect re-cooling, which only cools the system to low motional occupation of $\bar n\sim2$ phonons. While the \emph{M\o{}lmer-S\o{}rensen} (MS) gate is insensitive to the motional occupation to first order~\cite{Kirchmair2009}, higher phonon number leads to a larger sensitivity to calibration errors. 
Moreover, the recooling process takes \unit{3}{\milli\second}, during which the system experiences some dephasing. Hence, the recycling and non-recycling experiments are not directly comparable. It is, however, anticipated that the technical limitations can be overcome through the use of mid-circuit ground-state cooling and faster recycling schemes, such that comparable fidelities between the two methods can be achieved, as would also be required for realizing quantum error correction.

\cref{fig:noise} shows the results of the fidelity and TVD measurements for an increasing amount of noise on the $2\times 2$ cluster state in comparison to numerical simulations. 
We observe an increasing gap between TVD and upper bound from the root infidelity estimate (cf. \cref{eq:estimate tvd relation}) with the amount of noise in a fixed quantum circuit. 
These results indicate that output distributions of states subject to a significant amount of dephasing noise may still have a TVD well below the root infidelity. 
Comparing the experimental results with the simulations also allows us to deduce the natural noise floor in the experiment. 

\begin{figure}[t]
  \includegraphics{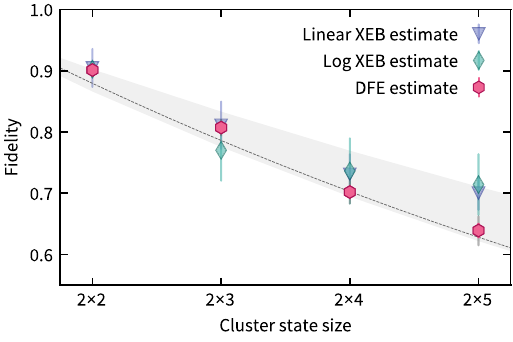}
  \caption{\label{fig:results average} 
  \textbf{Experimental results for average performance verification.}
  Average fidelity estimate from direct fidelity estimation (DFE) (pink hexagons), from linear XEB (triangles), and from logarithmic (log) XEB (diamonds, see Methods for the definition) using $1000$ random cluster states and $50$ shots per state. 
  Based on calibration data for the gate fidelities of single-qubit gates $f_{1Q} = 99.8\%$, two-qubit gates $f_{2Q} = 97.5 \pm 0.5\%$, and measurements $f_{M} = 99.85\%$, we compute a prediction for the fidelity (gray shaded line).
  We extract an effective local Pauli error probability of $1.7\%$ (dotted line), see \cref{app:average fidelity} of the SI. 
  Colored error bars represent the statistical $3\sigma$ error.
  For uncorrelated measurement noise, the fidelity estimate provides a lower bound to the true state fidelity. Gray error bars represent the worst-case systematic measurement error. 
  }
\end{figure}

We then measure the fidelity of cluster state preparations, averaged over the random circuits and show the results in \cref{fig:results average}.
We compare the fidelity estimates to the classical estimates of fidelity via XEB depending on the relative dimensions of the cluster, since in the circuit model the quality of XEB as a fidelity estimator depends on the circuit depth~\cite{ware_sharp_2023,morvan_phase_2023}.
We observe a consistently larger variance of the XEB estimate of the fidelity than of the direct fidelity estimate, and deviates for the $2\times5$ cluster. 
This may be due to the fact that the XEB fidelity depends on the type and strength of the experimental noise, but also the specific dimensions of the cluster state and the effective circuit ensemble~\cite{gao_limitations_2021,ware_sharp_2023,morvan_phase_2023}. 
Hence, while XEB generally seems to reflect the order of magnitude of the true fidelity, extreme care must be taken when using the XEB as an estimator of the fidelity.
\smallskip

\textbf{Discussion and conclusion.}
We conclude that direct (average) fidelity estimation provides an efficient and scalable means of certifying both single instances and the average quality of measurement-based computations.  
This is the case since the sample complexity of the fidelity estimate for arbitrary generalized stabilizer states is independent of the size of the system and the postprocessing is efficient.
Larger systems can therefore be verified with the same number of experiments as we have performed.

More generally, our results demonstrate that the measurement-based model of quantum computation provides a viable path toward efficient verification of quantum random sampling experiments, which is not known to be possible in the circuit model. 
In particular, all known methods for fidelity estimation \cite{flammia_direct_2011,huang_predicting_2020} in general scale exponentially with the number of qubits. We also note that, although MBQC is formally equivalent to the circuit model, relating a quantum circuit to an MBQC requires a space-time mapping and a feedforward procedure. Hence, our verification protocol at the level of the cluster state has no direct analog in circuit-based computations. 
While the experiments in this work are still far from the quantum advantage regime, we have successfully demonstrated how to use qubit recycling to perform large-scale MBQC with a qubit number that can be quadratically larger than the used ion register. 
This will enable trapped-ion quantum processors comprising on the order of 100 ions and depth 50 to achieve a fully verified quantum advantage in sampling from cluster states with more than $50\times 50$ nodes.

Besides trapped ions, several other platforms are compelling candidates for demonstrating a verifiable quantum advantage via random cluster state sampling. 
Examples include arrays of Rydberg atoms in optical tweezers, where the creation of large atom arrays~\cite{manetsch_tweezer_2024} has recently been demonstrated.
Another leading platform for cluster state generation is photonics~\cite{Istrati2020}, and continuous-variable optical systems where cluster states with up to 30 000 nodes have been experimentally prepared~\cite{Asavanant2019,Larsen2019}. 
Currently, these states are still Gaussian states and therefore not useful for quantum computing, but it is intriguing to think about how the non-standard topologies of continuous-variable cluster states might be exploited. Traditionally, such continuous variable systems have been used for boson sampling, rather than quantum circuit sampling. 
While boson sampling is not a universal model for computation, its efficient verification is possible for both photon-number \cite{chabaud_efficient_2021} and Gaussian \cite{aolita_reliable_2015} input states.
In practice, however, the verification measurements are entirely different in type compared to the sampling experiments, requiring a different apparatus. 
In contrast, for verifying MBQC states as performed in this work, the difference between sampling and verification is only local basis rotations. 
This makes MBQC a particularly compelling candidate for verifiable quantum random sampling.

\let\oldaddcontentsline\addcontentsline
\renewcommand{\addcontentsline}[3]{}
\section*{Methods}
\subsection*{Verification protocols}
\let\addcontentsline\oldaddcontentsline%


MBQC with cluster states is amenable to various types of verification. 
In particular, we can perform single-instance verification, that is, verification of a single quantum state using many copies of that state. 
We also perform average verification, that is, an assessment of the quality of state preparations averaged over the ensemble of measurement-based computations defined by the random choices of single-qubit rotation angles $\beta$.  
We distinguish classical means of verification in which we only make use of classical samples from the cluster state measured in a fixed (the Hadamard) basis, and quantum means of verification in which we measure the cluster state in various different bases. 

\paragraph*{Single-instance verification.}
In order to perform single-instance verification we apply direct fidelity estimation~\cite{flammia_direct_2011}, which uses single-qubit measurements on preparations of the target state $\ket \psi$.
Since the target state vector $\ket \psi$ for our random sampling problem is a locally rotated stabilizer state, with stabilizer operators $S_i$, $\rho = \ket \psi \bra \psi$ is the projector onto the joint $+1$-eigenspace of its $N$ stabilizers. 
We can therefore expand $\rho$ as the uniform superposition over the elements of its stabilizer group $\mc S = \langle S_1, \ldots, S_{N} \rangle$, where $\langle S_1, \ldots, S_{N} \rangle$ denotes the multiplicative group generated by $S_1, \ldots, S_N$. 
The fidelity can then be expressed as 
\begin{align}
  F &= \frac 1{2^{N}} \sum_{s \in \mc S} \langle s \rangle_\rho = \frac 1{2^{N}} \sum_{s \in \mc S} \sum_{\sigma = \pm 1} \langle\pi_s^\sigma \rangle_\rho \cdot \sigma,  
  \label{eq:dfe single}
\end{align}
where $s = \pi_s^+ - \pi_s^-$ is the eigendecomposition of the stabilizer $s$ into its $\pm 1$ subspaces, and $\langle \cdot \rangle_\rho = \tr[\rho \cdot]$ denotes the expectation value. This suggests a simple verification protocol where in each run a uniformly random element of $\mc S$ is measured on $\rho$. Averaging over the measurement outcomes $\sigma$ then gives an unbiased estimate of the fidelity according to \cref{eq:dfe single}.
Since the measurement outcomes $\sigma$ are bounded by $1$ in absolute value, we can estimate the average up to error $\epsilon$ using a number $M$ of  measurements from $\mc S$ that scales as $1/\epsilon^2$ and is independent of the number of qubits.

We also directly estimate the TVD between the empirical distribution and the ideal distribution. 
Note that estimating the TVD is sample-inefficient since the empirical probabilities need to be estimated, requiring exponentially many samples \cite{hangleiter_sample_2019}. 
It is also computationally inefficient since the ideal probabilities need to be computed. 

\paragraph*{Average-case verification.}
We measure the average quality of the cluster state preparations $\rho_\beta$ by their average state fidelity 
\begin{align}
  \label{eq:average state fidelity cluster methods}
  \overline F \coloneqq \Eb_{\beta}[\bra {\psi_\beta} \rho_\beta \ket {\psi_\beta}]
\end{align}
with the generalized cluster state $\ket {\psi_\beta}$ with random angles $\beta \in \{0, \frac \pi 4, \ldots, \frac{7\pi}4 \}^{n \times m }$.
Here, $\mb E_\beta[\cdot]$ denotes the expectation value over random $\beta \in [8]^{nm}$, where we let $[8] = \{1,2, \ldots, 8\}$ and $[k]^l = [k]\times \cdots \times [k]$ $l$ times. 

In order to classically estimate the average state fidelity, one can make use of \emph{cross-entropy benchmarking} (XEB) as proposed by \textcite{boixo_characterizing_2018,arute_quantum_2019_short}. 
XEB makes use of the classical samples from a distribution $Q$ and aims to measure how distinct $Q$ is from a target distribution $P$. 
The linear and logarithmic XEB fidelities between $Q$ and $P$ are defined as 
\begin{align}
   f_\lint(Q,P) & \coloneqq 2^n \sum_x Q(x)P(x) -1,  \label{eq:linear xeb methods}\\
    f_\logt(Q,P) & \coloneqq - \sum_x Q(x)\log P(x),  \label{eq:log xeb methods}
 \end{align}  
respectively.
Letting $P_\beta$ be the output distribution of $\ket{\psi_\beta} $ and $Q_\beta$ the output distribution of $\rho_\beta$ after Hadamard-basis measurements, we can estimate the average state fidelity from the average linear (logarithmic) XEB fidelity
\begin{align}
\overline f_{\lint \, (\logt)} &\coloneqq  \mb E_\beta \left[ f_{\lint \, (\logt)}(Q_\beta, P_\beta) \right], 
\label{eq:average xeb fidelity}
\end{align}
assuming that the total noise affecting the experimental state preparation $\rho_\beta$ is not correlated with $\proj{\psi_\beta}$. 
In order to estimate the (average) XEB fidelities, we need to compute the ideal output probabilities $P_\beta(x)$ and average those over the observed samples $x$. 
This renders the XEB fidelities a computationally inefficient estimator of the fidelity. 
They are \emph{sample-efficient} estimators, \cite{hangleiter_sampling_2021}, however, provided that the target distribution $P_\beta$ satisfies the expected exponential shape for deep random quantum circuits (or larger cluster states).
That is, to achieve an additive estimation error $\epsilon$, a polynomial number of samples in $n$ and $1/\epsilon$ are required. 
In \cref{app:average fidelity} of the SI, we provide the details of the estimation procedure.

To date, XEB is the only available means of practically verifying (on average or in the single-instance) universal random quantum circuits.

Here, we observe that in the measurement-based model of quantum computations \emph{fully efficient} (i.e., computationally and sample-efficiently) average-case verification is possible using single-qubit measurements. 
In fact, we observe that direct fidelity estimation can be extended to measure the average fidelity of random MBQC state preparations.
To this end, we observe that the average state fidelity \eqref{eq:average state fidelity cluster methods} can be expressed analogously to \cref{eq:dfe single} as 
\begin{align}
  \overline F  = \frac 1{2^{n  m}} \frac 1 {8^{n  m}} \sum_{\beta \in \frac\pi 4 \cdot [8]^{nm}} \sum_{s_\beta \in \mc S_\beta} \sum_{\sigma = \pm 1} \langle\pi_{s_\beta}^\sigma \rangle_{\rho_\beta} \cdot\sigma,  
\end{align}
where $\mc S_\beta$ denotes the stabilizer group of the locally rotated cluster state $\ket {\psi_\beta}$ with rotation angles $\beta$, $\pi_{s_\beta}^\sigma$ is the projector onto the $\sigma$-eigenspace of $s_\beta \in \mc S_\beta$, and .
Hence, in order to estimate the average state fidelity with respect to the choice of $\beta$, we simply need to sample uniformly random rotation angles $\beta$, and elements $s_\beta$ from the stabilizer group $\mc S_\beta$ and then measure $s_\beta$ on the state preparation $\rho_\beta$ of $\ket{\psi_\beta}$, yielding outcome $\sigma \in \{\pm1\}$. 
Averaging over those outcomes yields an estimator of the average state fidelity with the same sample complexity as direct fidelity estimation has for a single instance. 
As discussed below, the only assumption required to trust the validity of the result is that the noise in the local single-qubit measurements does not behave adversarially. 
Direct fidelity estimation and direct average fidelity estimation thus provide a unified method for efficiently assessing the single-instance quality and the average quality of MBQC state preparations.

\let\oldaddcontentsline\addcontentsline
\renewcommand{\addcontentsline}[3]{}
\subsection*{Finite sampling and error bars}
\let\addcontentsline\oldaddcontentsline%

When performing direct fidelity estimation of a fixed cluster state, the simplest protocol is to sample an element $s \in \mc S$ of the stabilizer group uniformly at random and measure $s$ once; cf.\ \cref{eq:dfe single}. 
In this case, the samples are distributed binomially with ideal probability $p = \sum_s \langle \pi_s^\sigma \rangle_\rho /2^N$, and the error on the mean estimation is given by the standard deviation of the observed binomial distribution. 
However, in practice, it is much cheaper to repeat a measurement of a stabilizer than to measure a new stabilizer, which requires a different measurement setting.
This is why we estimate the fidelity according to the following protocol. 
We sample $K$ stabilizers uniformly at random and measure each of them $M$ times, obtaining an empirical estimate of the conditional expectation value $\mb E[\sigma | s] = \sum_{\sigma = \pm1} \tr[\rho \pi_s^\sigma] \sigma$.
In \cref{app:mean error} of the SI, we show that the variance of the fidelity estimator $\hat F =(KM)^{-1} \sum_{i=1}^K \sum_{j=1}^M \sigma_{i,j}$, where $\sigma_{i,j}$ is the outcome of measuring stabilizer $s_i$ the $j^{\text{th}}$ time, is given by 
\begin{align}
  \var[\hat F]  = \frac 4 {KM}   (\mb E[p_s] (1 - \mb E[p_s])) + \frac 4 {K}\left( 1-\frac 1M \right) \var[p_s] .
  \label{eq:total variance}
\end{align}
Here, the expectation value and variance are taken over $s \in \mc S$ and $p_s = \tr[\rho \pi_s^{+1}]$, respectively. 
Furthermore, the same results carry over to the average fidelity estimate, since sampling from the stabilizer group $\mc S$ of a single cluster state is now replaced by sampling a random choice of angles $\beta$, and random element of the corresponding stabilizer group $\mc S_\beta$, not altering the variance.

\cref{eq:total variance} gives rise to an optimal choice of $K$ and $M$ for a fixed total number of shots $K\cdot M$, depending on the expectation value and variance of the stabilizer values $p_s$ and the experimental trade-off between repetitions of the same measurement and changing the measurement setting. 
In particular, if the distinct elements of the stabilizer group have a small variance over the imperfect state preparation $\rho$, a larger choice of $M$ might be advantageous. 
In practice, for the instances in which we have abundant data, we subsample the data in order to remain in the situation $M=1$ of \cref{eq:dfe single}, while in the case of sparse data, we make use of a larger number of shots $M$ per stabilizer. 

The variance of the estimate of the XEB fidelity is also given by the law of total variance, generalizing \cref{eq:total variance}, and spelled out in detail in \cref{app:mean error} of the SI. 
Finally, for the TVD, we estimate the error using bootstrapping by resampling given the observed distribution.  
Specifically, we repeatedly sample from the empirical distribution the same number of times as the experiment and compute the TVD of the samples to the sampled distribution. 
The resulting TVD follows a Gaussian distribution of which we show the $3\sigma$ interval estimated from $1000$ iterations.

\let\oldaddcontentsline\addcontentsline
\renewcommand{\addcontentsline}[3]{}
\subsection*{Measurement errors}
\let\addcontentsline\oldaddcontentsline%

A key assumption for the efficient verification of the cluster states prepared here is the availability of accurate,  well-characterized single-qubit measurements. 
A deviation in the measurement directly translates into a deviation in the fidelity estimate, and hence a high measurement error in the worst case translates into a high error in the resulting fidelity estimate. 
Because the single-qubit measurements we use comprise single-qubit gates followed by readout in a fixed basis, the measurement error has two main contributions: 
(i) imperfections in the single-qubit rotations for the basis choice, 
and (ii) imperfections in the readout.

The single-qubit gate errors are well characterized by randomized benchmarking, showing an average single-qubit Clifford error rate of $3\pm 2\cdot 10^{-4}$~\cite{Ringbauer2021} for the recycling device and $14\pm 1\cdot 10^{-4}$~\cite{Pogorelov2021} for the second device. 
The native $Z$ measurement is then performed by scattering photons on the short-lived $\mathrm{S}_{1/2}\leftrightarrow \mathrm{P}_{1/2}$ transition. Ions in the $\ket{0}$ state will scatter photons, while ions in the $\ket{1}$ state remain dark. Hence, there are two competing contributions to the readout error.
On the one hand, long measurement times suffer from amplitude damping noise due to spontaneous decay of the $\ket{1}$ state (lifetime $\sim 1.15$s) during readout. On the other hand, for short readout times, the Poisson distributions for the two outcomes will start to overlap, leading to discrimination errors. 
In the experiments presented here, the second contribution is suppressed to well below $10^{-5}$ by using measurement times of 1ms for the recycling device and 2ms on the non-recycling device, 
leaving only a spontaneous decay error of $< 1 \cdot 10^{-3}$~\cite{schindler_quantum_2013} for the recycling device and $< 2 \cdot 10^{-3}$ for the non-recycling device. 
Hence, the worst-case readout error is $< 1.5 \cdot 10^{-3}$ per qubit for the recycling device and $< 3.5 \cdot 10^{-3}$ per qubit for the second device. 
Given the single-qubit readout error $e_1$, the overall measurement error on an $n$-qubit device is then given by $e_M = 1-  (1- e_1)^n$. 

Given a true pre-measurement state fidelity $F$, we consider the effect of the measurement errors on the estimated fidelity $\hat F$.
In the one extreme case, the measurement errors flip the sign of the stabilizers with value $+1$ on the pre-measurement state, but keep the sign of those with a $-1$ outcome, resulting in a reduced state fidelity $\hat F_{\min} = 2 ((1 + F)/2 - e_M \cdot (1+ F)/2) -1 $. In the other extreme case, they flip the sign of only the stabilizers with value $-1$ on the pre-measurement state yielding $\hat F_{\max} = 2 ((1 + F)/2 + e_M \cdot (1- F)/2) -1 $. This defines the worst-case error interval for $\hat F$ as $[(\hat F - e_M)/(1- e_M), (\hat F + e_M)/(1- e_M)]$.

If on the other hand the measurement errors are benign, i.e., uncorrelated from the circuit errors, they will flip all stabilizers regardless of their value on the pre-measurement state with equal probability. 
In this case, the measured fidelity satisfies $\hat F = F \cdot (1- 2 e_M)$ so that we can deduce the true fidelity $F$ from the measured fidelity and the measurement error. 
Note that in this case, the measured state fidelity is always a lower bound to the true state fidelity.

\let\oldaddcontentsline\addcontentsline
\renewcommand{\addcontentsline}[3]{}
\subsection*{Noisy circuits}\label{ssec:noisy circuits}
\let\addcontentsline\oldaddcontentsline%

\begin{figure}
  \includegraphics[width=\linewidth]{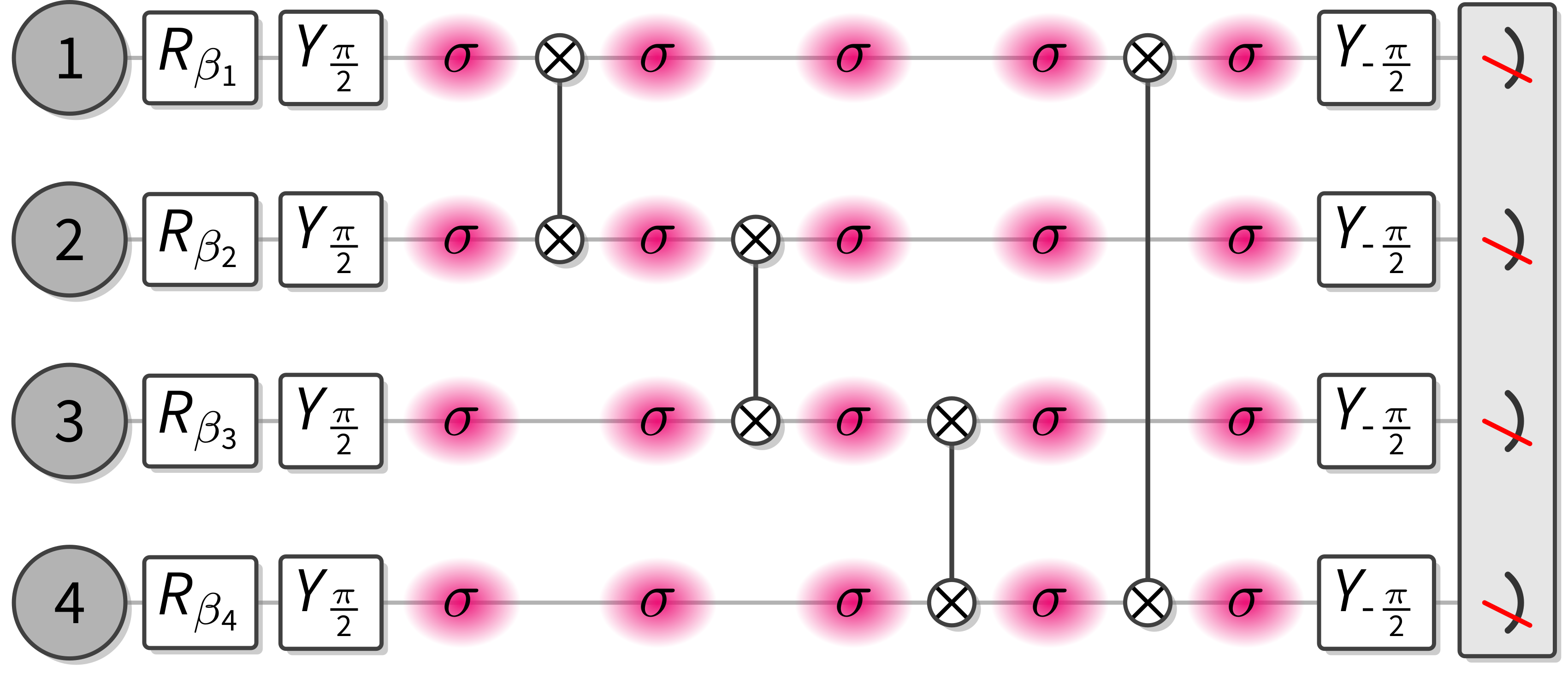}
  \caption{\label{fig:noisy circuit} \textbf{Noisy circuits for the $2\times 2$ cluster.} Dephasing noise is simulated by adding random (virtual) $Z$ rotations on all qubits after the initial state preparation and after each MS gate, see Methods. This amounts to roughly equidistant time steps. 
  The rotation angles for the $Z$ rotations are drawn randomly from a normal distribution with zero mean and standard deviation $\sigma \in [0,0.2\pi]$ every 50 shots. 
  For correlated noise, the parameters in each time step are chosen equally and for uncorrelated noise, they are chosen independently.}.
\end{figure}

In order to study the influence of experimental noise on the reliability and tightness of our bounds on the TVD, we artificially induce dephasing noise on the $2\times 2$ cluster. This simulates a reduced spin-coherence time, which could come from laser phase noise or magnetic field noise. These are the dominant error sources in the experiment. To this end, we pick a fixed instance of the $2\times 2$ cluster and add small random $Z$ rotations on all qubits at roughly equidistant time steps. Specifically, we apply virtual $Z$ gates (i.e., realized in software as an appropriate phase shift on all subsequent gate operations) after the initial local state preparation gates, and again after each MS gate, see \cref{fig:noisy circuit}. In each run of the experiment (with 50 shots each), we randomly pick rotation angles for the virtual $Z$ gates from a normal distribution with 0 mean and standard deviation $\sigma$. Here $\sigma$ is a measure of the noise strength and corresponds to a local phase-flip probability of $\xi/2$, where $\xi=1-e^{-\sigma^2/2}$. If we want to engineer global, correlated noise, we use the same angle for all $Z$ gates in a given ``time-step'', whereas for engineering local, uncorrelated noise we pick each angle independently. We then average these random choices over 50 instances for the fidelity estimate and 150 instances for the TVD. This averaging turns the random phase shifts into independent (correlated) dephasing channels in the case of local (global) noise. 
This effectively appears as single-qubit depolarizing noise after every two-qubit gate with a local Pauli error probability of $3\gamma/4$, where $\gamma=1-e^{-0.310\sigma^2}$, where the constant was obtained from a numerical fit to simulated data, see \cref{app:average fidelity} of the SI for details. 
\medskip

\textbf{Acknowledgements.} 
D.H.\ acknowledges funding from the U.S.\ Department of Defense through a QuICS Hartree fellowship. 
This work was completed while D.~H.\ was visiting the Simons Institute for the Theory of Computing.
The Berlin team acknowledges funding from the BMBF (DAQC, MUNIQC-ATOMS), the DFG (specifically EI 519/21-1 on paradigmatic quantum devices, but also CRC 183), the Einstein Foundation (Einstein Research Unit), the BMWi (EniQmA, PlanQK) and the Studienstiftung des Deutschen Volkes.
This research is also part of the Munich Quantum Valley (K-8), which is supported by the Bavarian state government with funds from the 
Hightech Agenda Bayern Plus. It has also received funding from the EU's Horizon 2020 research and innovation programme under the Quantum Flagship projects PASQuanS2 and Millenion. JBV acknowledges funding from EU Horizon 2020, Marie Skłodowska-Curie GA.~Nr.~754446 - UGR Research and Knowledge Transfer Fund Athenea3i; Digital Horizon Europe project FoQaCiA,~GA.~
Nr. 101070558., and  FEDER/Junta de Andalucía program A.FQM.752.UGR20. The Innsbruck team acknowledges support by the Austrian Science Fund (FWF), through the SFB BeyondC (FWF Project No.~F7109) and the EU QuantERA project T-NiSQ (I-6001), and the Institut f\"ur Quanteninformation GmbH. We also acknowledge funding from the EU H2020-FETFLAG-2018-03 under Grant Agreement No.~820495, by the Office of the Director of National Intelligence (ODNI), Intelligence Advanced Research Projects Activity (IARPA), via US Army Research Office (ARO) grant No.~W911NF-20-1-0007, and the US Air Force Office of Scientific Research (AFOSR) via IOE Grant No. FA9550-19-1-7044 LASCEM. This research was funded by the European Union under Horizon Europe Programme---Grant Agreement 101080086---NeQST. Funded by the European Union (ERC, QUDITS, 101039522). Views and opinions expressed are, however, those of the author(s) only and do not necessarily reflect those of the European Union or the European Commission. Neither the European Union nor the granting authority can be held responsible for them.\smallskip

\textbf{Author contributions.} 
M.R.\, T.F.\, and L.P.\ performed the experiments. 
D.H., M.H., and M.R.\ analyzed the data and performed the numerical simulations. 
D.H.\ and M.H.\ derived the theory of average (XEB) fidelity in MBQC. 
D.H., P.K.F.,  M.H., J.B-V., and J.E.\ conceived of the protocols. 
M.R., T.F., C.E., C.D.M., R.S., M.M., I.P., L.P., P.S., and T.M.\ contributed to the experimental setup.
R.B., J.E., and T.M.\ supervised the project.
D.H.\ and M.R.\ wrote the initial draft of the manuscript. 
All authors contributed to discussions and writing the final manuscript. \smallskip

\textbf{Competing interests.} The authors declare no competing interests. 

\textbf{Data availability.} The experimental data and python code (which uses the Qiskit package \cite{Qiskit}) used in the numerical simulations and data analysis can be obtained from the authors.

\let\oldaddcontentsline\addcontentsline
\renewcommand{\addcontentsline}[3]{}
\putbib
\let\addcontentsline\oldaddcontentsline
\end{bibunit}

\begin{bibunit}
  
\onecolumngrid
\cleardoublepage
\thispagestyle{empty}

\setcounter{page}{1}
\setcounter{equation}{0}
\setcounter{footnote}{0}
\setcounter{section}{0}
\setcounter{figure}{0}
\makeatletter

\renewcommand{\thepage}{\@roman\c@page}
\pagenumbering{roman}
\renewcommand{\thesection}{S\@arabic\c@section}
\renewcommand{\theequation}{S\@arabic\c@equation}
\renewcommand{\thefigure}{S\@arabic\c@figure}
\renewcommand{\thetable}{S\@arabic\c@table}
\makeatother
\begin{center}
\textbf{\large Supplementary Information for \\``Verifiable measurement-based quantum random sampling with trapped ions''}\\
\vspace{2ex}
\end{center}

\tableofcontents
%

\section{Fidelity witness}
\label{app:witness}

In addition to the fidelity estimate, we have also measured a witness for the fidelity on the generalized cluster states \cite{cramer_efficient_2010,hangleiter_direct_2017}. 
A fidelity witness of a quantum state $\ket \psi$ is a Hermitian operator $W$ with the properties
\begin{align}
  \tr[\rho W] =1 \Leftrightarrow \rho = \proj \psi \quad \text{ and } \quad \tr [\rho W] \leq F(\rho, \ket \psi). 
\end{align}
In other words, the expectation value of a fidelity witness provides a meaningful lower bound on the fidelity of a state preparation $\rho$ with a target state $\ket \psi$.

\begin{figure}
  \includegraphics{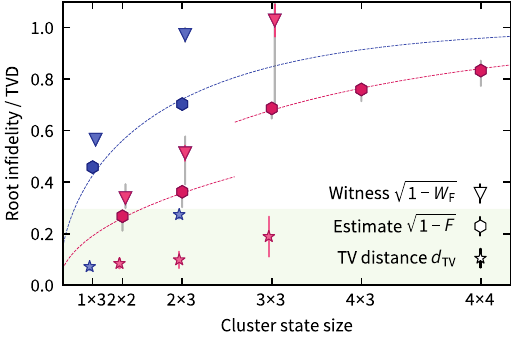}
  \caption{\label{fig:results + witness} 
  \textbf{Experimental results for cluster states with and without recycling.}
  (\cref{fig:results} from the main text including the fidelity witness)
  Root infidelity witness $\sqrt{1- W_F}$ (triangles), root infidelity estimate $\sqrt{1- F}$ (hexagons), and empirical total-variation distance (stars) for single instances of random MBQC cluster states with recycling (blue) and without (pink). Colored error bars represent the 3$\sigma$ interval of the statistical error.
  For uncorrelated measurement noise, the systematic measurement error only reduce the measured state fidelity. 
  The worst-case behaviour of the measurement error is represented by the gray error bars. 
  Modelling the circuit noise as local depolarizing noise after each entangling gate (dotted lines), we extract effective local depolarizing error probabilities  of 5.3\%, 2.6 \% and 1\% for the cluster with recycling, the cluster in the large register, and the cluster in the small register, respectively; see \cref{eq:dalzell formula} below for details. 
  } 
\end{figure}

To construct a fidelity witness for the cluster states, we observe that the pre-measurement cluster state in the protocol is the ground state of a commuting, local Hamiltonian $H$ with gap $\Delta = 2$, consisting of local terms given by the locally rotated stabilizers $S_i$ of the cluster state. 
The energy of the local terms in an imperfect state preparation $\rho$ yields a certificate for the fidelity $F = \bra \psi \rho \ket \psi $ with the target state vector $\ket{\psi}$ in terms of the witness $W_F$ as~\cite{cramer_efficient_2010,hangleiter_direct_2017}
\begin{align}
\label{eq:witness}
  W_F = 1 - \frac 12  \sum_{i = 1}^{N} \langle S_i \rangle_\rho  \leq   F. 
\end{align}
In particular, this implies that the root infidelity bound \eqref{eq:estimate tvd relation} can be supplemented as 
\begin{align}
\label{eq:estimate witness tvd relation}
  d_{\text{TV}} \leq \sqrt{ 1- F} \leq \sqrt{1-W_F}. 
\end{align}

We have measured the fidelity witness for the same quantum states prepared for \cref{fig:results} in the main text by measuring their stabilizers, see \cref{fig:results + witness}. We find that the upper bound \eqref{eq:estimate witness tvd relation} remains meaningful (i.e., smaller than one) only for very small sizes of the cluster, and conclude that the fidelity witness has very limited use in the presence of a significant amount of noise in the system.

\section{Experimental details}
\label{app:experimental details}

In \cref{tab:num samples}, we detail the number of experimental shots for sampling and verification used in \cref{fig:results} and \cref{fig:results + witness}. 

\begin{table}
  \begin{tabular}{r l >{\ } l <{\ }l>{\ }l >{\ }l}
  \toprule
  & Size & Sampling shots & \multicolumn{3}{c}{Verification shots}\\
  &  &  & $M(W_F)$ & $K(F)$ & $M(F)$\\
  \midrule 
  \midrule 
    Recycling & 1 x 3& 15 000 & 15 000 & 118 980 & 1\\
    & 2 x 3 & 62 500 & 3 600 & 223 980& 1\\
  \midrule 
    Non-recycling & 2 x 2 & 15 000 & 10 000 & 157 611& 1 \\
    & 2 x 3 & 15 000 & 2 500 & 154 177 & 1  \\
    & 3 x 3 & 15 000 & 500 & 225 865&  1 \\ 
    & 4 x 3 & --- 
     & 50 & 110 408 & 1 \\
    & 4 x 4 & --- 
    & 50 & 640 & 50 \\
  \bottomrule
  \end{tabular}
  \caption{\label{tab:num samples} 
  \textbf{Number of samples used in \cref{fig:results}.}
  $M(W_F)$: Number of shots per stabilizer for the witness. 
  $K(F)$: Number of randomly drawn stabilizers for the fidelity estimate.
  $M(F)$: Number of shots per stabilizer for the fidelity estimate. 
  }
\end{table}

\section{Compiled circuits}
\label{app:compiled circuits}
The quantum circuit giving rise to the cluster state can be succinctly written as 
\begin{align}
  \ket{\text{CS}(\beta)} = \left(\prod_{\langle k,l\rangle} CZ_{k,l}\right) \left( \prod_{k} \ee^{-\ii \beta_{k} Z_k }H_k \right) \ket{0^N}.
  \label{eq:SuppCS}
\end{align}
Here $\langle k,l\rangle$ denotes all pairs of neighboring ions in the respective cluster state.
We express this circuit in terms of the native gates in the ion-trap architecture, pairwise addressed M{\o}lmer S{\o}rensen gates $\ms_{k,l}$ and arbitrary rotations around an axis in the X-Y plane $R(\theta, \phi)$, 
given by 
\begin{align}
   \ms_{k,l} &= \exp\left(-\ii \frac \pi 4 X_k X_l\right),\\
   R(\theta, \phi) &= \exp\left( -\ii \frac \theta 2 \left(\cos(\phi) X + \sin(\phi ) Y\right)\right) = \begin{pmatrix} \cos\frac \theta 2& - \ii \ee^{-\ii \phi} \sin\frac\theta 2\\
   \ii\ee^{\ii \phi} \sin\frac \theta 2 & \cos\frac \theta 2\end{pmatrix}.
 \end{align} 
Here, the polar angle $\theta$ corresponds to the laser pulse area, while the azimuthal angle $\phi$ is determined by the phase of a laser pulse. 
With these definitions, we observe some properties of the rotation gate
\begin{itemize}
   \item $X(\theta) = \ee^{- \ii \frac \theta 2 X} = R(\theta,0)$,
   \item $Y(\theta) = R(\theta, \pi/2)$,
   \item $R(\theta, \phi + \pi)\ket 0  = \ee^{\ii \pi/2} Z R(\theta, \phi) \ket 0 $.
 \end{itemize}
Since the phase can be controlled much more precisely than the pulse area, it is advantageous to only perform $\theta = \pi/2$-pulses with variable azimuthal angle $\phi$. 
Now, let us decompose $CZ$ in terms of the above gates. 
We start by writing 
\begin{align}
  CZ_{k,l} = \ee^{-\ii \frac \pi 4} \cdot \ee^{\ii \frac \pi 4 Z_i}\ee^{\ii \frac \pi 4 Z_j}  \ee^{ - \ii \frac \pi 4 Z_i Z_j } , 
\end{align}
and then observe that $Y(\pi/2) Z(\pi) = H$, and since the rotation angle of $Z$ does not matter for computational-basis measurements, we can replace $H$ by $Y(\pi/2)$ and $H^\dagger = Z(-\pi) Y(-\pi/2)$. 
Hence $Y(-\pi/2) X(\theta) Y(\pi/2) = Z(\theta)$ such that
\begin{align}
  CZ_{k,l} = \ee^{-\ii \pi/4} \left(\prod_{i = k,l}Y_i(-\frac \pi 2)X_i(-\frac \pi 2)\right) 
  \ms_{k,l} \left(\prod_{i = k,l}Y_i(\frac \pi 2)\right). 
\end{align}
Putting this back into the first part of Eq.~\eqref{eq:SuppCS} we observe that intermediate $Y(\pm \pi/2)$ gates cancel and $X$-gates commute to the left leaving 
\begin{align}
  \prod_{\langle k,l \rangle} CZ_{k,l}= \ee^{-\ii M \pi/4} \left(\prod_{i = 1}^N Y_i\left(-\frac \pi 2\right)X_i\left(-\frac {\deg(i) \pi} 2\right)\right) \left(\prod_{\langle k,l \rangle} \ms_{k,l} \right) \left(\prod_{i = 1}^N Y_i\left(\frac \pi 2 \right)\right), 
\end{align}
where $M$ is the total number of MS-gates and $\deg(i)$ is the degree of site $i$ (i.e., the number of links in the cluster).
We can further simplify
\begin{align}
  Z(\beta) H \ket 0 =Z(\beta +\pi) Y(\pi/2) \ket 0 = \ee^{- \ii \frac \beta 2} R(\pi/2, \beta + \pi/2) \ket 0  = \ee^{- \ii (\frac \beta 2 - \frac\pi2)} Z(\pi) R(\pi/2, \beta - \pi/2) \ket 0, 
\end{align}
and since $Y(\pi/2)Z(\pi) = X(\pi)Y(\pi/2)$ we obtain the total circuit 
\begin{align}
  \ket{\text{CS}(\beta)} &= \ee^{-\ii (M \pi/4 + \frac 12\sum_i \beta_i- N \pi/2)} \left(\prod_{i = 1}^N Y_i\left(-\frac \pi 2\right)X_i\left(-\frac {(\deg(i) - 2) \pi} 2\right)\right) \left(\prod_{\langle k,l \rangle} \ms_{k,l} \right) \left(\prod_{i = 1}^N Y_i\left(\frac \pi 2 \right) R_i\left(\frac \pi 2 , \beta - \frac \pi 2  \right) \right)\\
 & = \ee^{-\ii (M \pi/4 + \frac 12\sum_i \beta_i- N \pi/2)} \left(\prod_{i = 1}^N Y_i\left(-\frac \pi 2\right)\left(X_i\left(-\frac {\pi} 2\right) \right)^{\deg(i) - 2 } \right) \left(\prod_{\langle k,l \rangle} \ms_{k,l} \right) \left(\prod_{i = 1}^N Y_i\left(\frac \pi 2 \right) R_i\left(\frac \pi 2 , \beta - \frac \pi 2  \right) \right). \nonumber
\end{align}
To perform a measurement in the Hadamard basis, we can now rotate back using $H = Z Y(-\pi/2)$, which can be absorbed in the leftmost layer of $Y(-\pi/2)$-gates, to give a layer of $Y(-\pi) = Y$ gates. Since these are just a phase flip and a bit flip, we can leave them out in the experiment and perform them in the classical postprocessing.

\section{Computing the threshold fidelity from Stockmeyer's argument}
\label{app:stockmeyer argument}

In the rigorous argument for the hardness of quantum random sampling, one makes use of Stockmeyer's algorithm \cite{stockmeyer_complexity_1983}, an algorithm in the third level of the polynomial hierarchy, in order to estimate \#P-hard probabilities; see Chapter 2 of Ref.~\cite{hangleiter_sampling_2021} for an accessible explanation of the algorithm. 
Let us briefly summarize the argument here, and refer the reader to 
Ref.~\cite{hangleiter_computational_2023} for a more detailed exposition. 
We assume that there exists a classical algorithm $\mathcal{A}$ that samples from the output distribution $p_U$ with probabilities $p_U(x) = | \bra x U \ket 0|^2$ up to an additive total-variation-distance error $\varepsilon$. 
We then feed $\mathcal{A}$ into Stockmeyer's algorithm and ask that it compute an approximation $q_U(x)$ of the probability $p_U(x)$ for some binary string $x$. 

The crucial step in the hardness proof then consists in balancing the error stemming from Stockmeyer's algorithm itself and the error incurred from the assumption to obtain a multiplicative approximation up to a factor $1/4$ with constant probability over the choice of $U$ and $x$.
The relevant expression is given by applying Markov's inequality yielding that with probability $1- \delta$
\begin{equation}
  |p_U(x) - q_U(x) | \leq \frac{p_U(x)}{\poly(n)} + \frac{\varepsilon}{2^n \delta} \left( 1 + \frac{1}{\poly(n)} \right) . 
\end{equation}
We conjecture the distribution $p_U$ to anticoncentrate in the sense that
\begin{equation}
  \mathrm{Pr}_U\left[ p_U(x) \geq \frac{1}{2^n}\right] \geq \gamma , 
\end{equation}
for some constant $\gamma > 0 $. 
As a result we obtain that with probability $\gamma ( 1- \delta)$ Stockmeyer's algorithm yields a relative-error $\varepsilon/\delta + o(1)$ approximation of $p_U(x)$.  
Hence, assuming that any $\gamma ( 1- \delta) $-fraction of the instances is \#P-hard to approximate up to relative error $\varepsilon/\delta + o(1)$, then this argument shows that one can approximate \#P-hard quantities in the third level of the polynomial hierarchy -- counter the established belief in theoretical computer science. 

Consequently, we can trade the fraction of instances we conjecture to be hard with the tolerated error $\varepsilon$ of the classical algorithm.
Making a bolder average-case conjecture results in a larger error bound from the argument. 
As discussed in Ref.~\cite{bermejo-vega_architectures_2018}, we numerically find $\gamma = 1/\mathrm{e}$. 
Setting the average-case hard fraction $\nu = 10^{-3}$, and relative-error $\min(1-1/\sqrt{2},\sqrt{2} - 1) =  0.2928$ approximation~\cite{fujii_commuting_2017}, we obtain a threshold total variation-distance $\nu/\gamma$ and consequently threshold  infidelity
\begin{align}
  1- F_T = (0.2928 \cdot (1- \nu / \gamma))^2 = 0.0857 .  
\end{align}
To summarize, we conjecture that any $10^{-3}$ fraction of the output probability instances is \#P-hard to approximate. 
We then obtain that sampling is hard up to total-variation distance $0.29$ with probability at least 1\%. Correspondingly, the threshold infidelity of accepting a quantum state is roughly 8.6\%; see \cref{fig:results}. 

We emphasize, however, that the threshold fidelity from the hardness argument above does not have a fundamental meaning in the sense that it does not imply easiness above the threshold. 
Rather, it gives a threshold such that---if the complexity-theoretic conjectures underlying the argument are true---hardness of sampling is guaranteed if the infidelity remains below the threshold. 
More generally, the takeaway should be that sampling from a family of distributions whose TVD to the target distributions does not increase if the cluster size is increased is likely a classically hard task; potentially even for a slow (polynomial) increase, see also Refs.\ \cite{aaronson_complexity-theoretic_2017,barak_spoofing_2021}. 

\section{Properties of the average fidelity in measurement-based quantum computing}
\label{app:average fidelity}

In this section, we discuss the average fidelity 
\begin{align}
  \label{eq:average state fidelity cluster}
  \overline F = \Eb_{\beta}[\bra {\psi_\beta} \rho_\beta \ket {\psi_\beta}]
\end{align}
of state preparations $\rho_\beta$ of the generalized cluster state $\ket {\psi_\beta}$ with respect to random local rotation angles $\beta \in [8]^{n \times m }\cdot \pi/4$.

First, in \cref{ssec:average dfe}, we show that the \emph{direct fidelity estimation} (DFE) protocol of \textcite{flammia_direct_2011} can be directly applied to estimating average state fidelities.
Then, in \cref{ssec:average xeb} we study properties of the cross-entropy measures in measurement-based computing, and in particular, discuss  some specifics of measurement-based quantum computing, in particular, the role of logical and physical circuits for these cross-entropy measures. 
Given this, we show in \cref{ssec:average_fid_from_xeb} how to use cross-entropy measures as an alternative way to estimate average fidelities, analogous to the theory of \emph{cross-entropy benchmarking} (XEB) for random circuits by \textcite{arute_quantum_2019_short}. 
We show analytically that under certain assumptions on the noise in the device, such cross-entropy measures can indeed also be applied in the context of MBQC to estimate average fidelities. 
Lastly, in \cref{ssec:xeb numerics} we support our analytical considerations with a numerical study comparing the resulting average fidelities from the DFE protocol with those obtained via XEB.

\subsection{Direct average fidelity estimation}
\label{ssec:average dfe}

In this section, we show how DFE can be directly applied to estimating the average cluster state fidelity $\overline F$. 
To this end, we observe that, analogously to the single-instance case discussed in the Methods section of the main text, 
\begin{align}
  \overline F &= \Eb_{\beta} \left [ \frac 1{2^{N}} \sum_{s_\beta \in \mc S_\beta} \langle s_\beta \rangle_{\rho_\beta} \right]\nonumber\\
  & = \frac 1{2^{N}} \frac 1 {8^{N}} \sum_{\beta \in \frac\pi 4 \cdot [8]^{N} } \sum_{s_\beta \in \mc S_\beta} \sum_{\sigma = \pm 1} \sigma \cdot\langle\pi_{s_\beta}^\sigma \rangle_{\rho_\beta}.  
  \label{eq:average dfe}
\end{align}
Here, the stabilizer group $\mc S_\beta = \langle S^\beta_1, \ldots, S^\beta_{N} \rangle$ and $\pi_{s_\beta}^\sigma$ is the projector onto the $\sigma$-eigenspace of $s_\beta \in \mc S_\beta$.
Hence, we obtain an unbiased estimator $\hat{\overline{F}}$ of the average fidelity by drawing a uniformly random pair $(\beta, S_\beta)$, measuring $S_\beta$ on $\rho_\beta$, and averaging over the measurement outcomes $\sigma$.

Since the measurement outcomes are bounded by $1$ in absolute value, we can estimate the average up to error $\epsilon$ using a number $M$ of uniformly random samples $(\beta, S_\beta)$ that scales as $1/\epsilon^2$ and is in particular independent of the number of qubits. 
As before, if we measure $M$ shots per random pair of state and stabilizer of which we draw $K$ many, the variance of this will be given by the variance formula 
\begin{align}
  \var[\hat {\overline F}] = \frac 4 {KM}   (\Eb_i[p_i] (1 - \Eb_i[p_i])) + \frac 4 {K}\left( 1-\frac 1M \right) \var_i[p_i] , 
\end{align}
where the expectation runs over the choice of random state and stabilizer as labeled by $i$. 

\subsection{Cross-entropy benchmarking of measurement-based quantum computing}
\label{ssec:average xeb}

\emph{Cross-entropy benchmarking} (XEB) has been proposed as a way to measure the average fidelity of quantum state preparations and has been applied to computations in the circuit model \cite{arute_quantum_2019_short,zhu_quantum_2022_short}, as well as to certain analog quantum simulations \cite{choi_emergent_2021}.
This benchmark has been developed by \textcite{boixo_characterizing_2018,arute_quantum_2019_short}, and the key property that makes it useful to experiments is that it can be sample-efficiently estimated for sufficiently random ensembles of circuits whose output probabilities are exponentially distributed. 
XEB has a complexity-theoretic interpretation in terms of the task dubbed \emph{heavy outcome generation} \cite{aaronson_complexity-theoretic_2017,aaronson_classical_2020}, 
an interpretation
as a proxy for the TVD under assumptions on the noise in the classical output distribution \cite{bouland_complexity_2019},
and provides a means to estimate the (average) quantum fidelity of the state from which the classical samples are produced \cite{arute_quantum_2019_short}, see Ref.~\cite{hangleiter_computational_2023} for an overview.

The most important XEB quantities are the so-called \emph{linear} and 
\emph{logarithmic XEB fidelities} (see, e.g.,
Ref.~\cite{hangleiter_computational_2023},
Section~V.B). 
For an ideal target probability distribution $P$ and a noisy distribution $Q$ on $n$-bit strings, we define those as 
\begin{align}
   f_\lint(Q,P) & \coloneqq 2^n \sum_x Q(x)P(x) -1, \label{eq:linear xeb}\\
   f_\logt(Q,P) & \coloneqq - \sum_x Q(x) \log P(x), \label{eq:log xeb}
 \end{align} 
respectively. 
These quantities can be empirically estimated by drawing samples $x_1,\dots,x_k$ from the noisy distribution $Q$ and averaging $P(x_1),\dots,P(x_k)$ or the logarithms of these, respectively.

For any single fixed circuit instance, the value of the XEB fidelities defined above and the actual quantum state fidelity of the state preparation before measurement could behave quite differently. A relation between XEB fidelity and quantum state fidelity can only be established on average over wide ensembles of circuits and only under additional assumptions described in more detail below. To this effect, we consider the average XEB fidelities $\overline f_\lint$ and $
 \overline f_\logt$, 
where the average is taken over the ensembles of circuit instances. As described in \Cref{ssec:average_fid_from_xeb}, those average XEB fidelities will serve us as estimators of the average
state fidelity $\overline{F}$.

\begin{figure}
  \includegraphics[width=\linewidth]{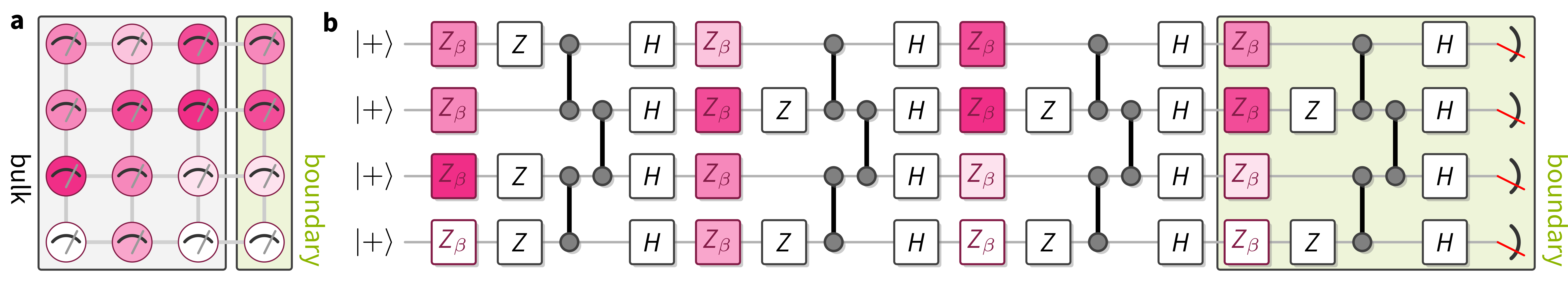}
  \caption{\label{fig:physical logical circuits}
  \textbf{Distinction between physical and logical circuit.} 
  \textbf{(a)} The \emph{physical circuit} is the circuit which we apply in the experiment. It is a constant-depth circuit on $n \times m$ qubits that comprises a layer of single-qubit Hadamard gates and local rotations, and a layer of entangling $CZ$ gates, followed by a measurement in the Hadamard basis. 
  We further distinguish by convention the \emph{bulk} of the lattice comprising the first $m-1$ columns and its \emph{last column}. 
  \textbf{(b)} The \emph{logical circuit} is the effective depth-$m$ circuit applied to the last column of qubits. It is generated by the measurements in the bulk of the cluster and, for each column of the physical state, comprises three layers: a layer of single-qubit rotations $Z_\beta$ with rotation angles $\beta$ corresponding to the respective position in the cluster and randomly applied $Z$ gates depending on whether or not the outcome of the corresponding qubit was $1$ or $0$, a layer of entangling $CZ$ gates, and a layer of Hadamard gates. 
  }
\end{figure}

 We note that the above notions of XEB quantities apply to any model of quantum computation. In the following, we will discuss aspects of XEB that are particular to the measurement-based quantum computing setting considered in this work. We will denote the output probability distributions of an ideal cluster state preparation $\ket{\psi_\beta} $ and an imperfect cluster state preparation $\rho_\beta$  upon measurement in the Hadamard basis, by $P_\beta$ and $Q_\beta$, respectively. Now, in contrast to the circuit model, in the context of MBQC, we seem to face a choice regarding which distributions to base our XEB quantities on: 
Recall that in MBQC, the measurement outcomes in the `bulk' of the system---that is, all qubits but those in the last (rightmost) column of the cluster---determine a \emph{logical circuit} applied to the state $\ket {+}^{\otimes n}$ on the `boundary' of the cluster---precisely that last column---see \cref{fig:physical logical circuits}.
According to this distinction between bulk and boundary, we split full outcome strings as $x = (x_b,x_f)$ into bulk outcomes $x_b \in \{0,1\}^{n(m-1)}$ and the outcomes on the final column $x_f \in \{0,1\}^n$.

Now, there are two output distributions that one could consider as inputs to XEB:  
First, there is the (joint) output distribution $P_\beta(x)=P_\beta(x_b,x_f)$ over the outcomes of the whole \textit{physical circuit}, that is, the circuit which we actually apply in the lab, including all $N=n\cdot m$ measurement outcomes. Alternatively, there is the output distribution $\tilde{P}_{\beta, x_b}(x_f)$ of the logical circuit over the $n$ outcomes $x_f$ of only the last column of qubits. The logical circuit is determined by the choices of angles $\beta$ as well as the $N-n$ measurement outcomes in the bulk $x_b.$ Hence, the logical outcome distribution is simply the conditional distribution $\tilde{P}_{\beta, x_b}=P_\beta(\cdot|x_b)$ and the noisy samples are distributed according to $\tilde{Q}_{\beta, x_b} =Q_\beta( \cdot |x_b)$. This suggests, that we could also consider \textit{logical} XEB quantities associated with the logical distribution $\tilde{P}_{\beta, x_b}(x_f)$ as follows
\begin{align}
  \tilde  f_\lint(\tilde{Q}_{\beta, x_b} ,\tilde{P}_{\beta, x_b}) & = 2^n \sum_{x_f} \tilde{Q}_{\beta, x_b}(x_f)\tilde{P}_{\beta, x_b}(x_f) -1, \\
   \tilde f_\logt(\tilde{Q}_{\beta, x_b},\tilde{P}_{\beta, x_b}) & = - \sum_{x_f} \tilde{Q}_{\beta, x_b}(x_f) \log \tilde{P}_{\beta, x_b}(x_f).
 \end{align}

A motivation for considering logical output distributions of an $n \times m $ cluster state is that, in contrast to the physical output distributions, the logical ones behave analogously to the outcome distribution of random circuits on $n$ qubits with depth $O(m)$. In particular, their statistical properties---which are an important ingredient in XEB theory---match up to constant factors \cite{haferkamp_closing_2020}. 
Conversely, if our goal is to estimate the average fidelity of the full cluster state, the XEB fidelity of the physical output distribution $P_\beta(x_b,x_f)$ seems to be the relevant quantity.  

As it turns out, however, the XEB fidelities associated with  physical and logical output distributions are in fact equivalent when used for average fidelity estimation via XEB. 
To see this, we have to consider the averaged XEB quantities introduced above. 
Concretely, we consider averages over the ensemble of circuits induced by a uniformly random choice of $\beta$ from $\frac \pi 4 \cdot [8]^{n \times m }$. 
This random choice induces via the map $\beta \mapsto P_\beta$, a distribution over ideal output distributions which we denote by $\mathcal{P}$. 
Similarly, we denote by $\mathcal{Q}$ the distribution over noisy output distributions induced via $\beta \mapsto Q_\beta$. 
Hence, the average XEB fidelities are written as 
\begin{align}
  \overline f_\lint (\mc Q, \mc P ) & = \Eb_{Q \sim \mc Q, P \sim \mc P } \left [  f_\lint(Q,P) \right],\\
  \overline f_\logt (\mc Q, \mc P ) & = \Eb_{Q \sim \mc Q, P \sim \mc P } \left [  f_\logt(Q,P) \right].
\end{align}
Analogously, the average logical XEB fidelities arise from drawing logical circuit instances $(\beta, x_b)$ according to the uniform choice of $\beta$ and $x_b$ from the marginal distribution with probabilities $Q_\beta(x_b) = \sum_{x_f} Q_\beta(x_f, x_b)$. 

Now, to relate the average XEB fidelites corresponding to physical and logical circuits, we will use the fact that, under the ideal distribution $P_\beta(x)$, the outcomes $x_b$ are uniformly distributed \cite{childs_unified_2005} so that $P_\beta(x_f, x_b) = 2^{-n(m-1)} P_\beta(x_f|x_b)$. 
 Then, we compute
\begin{align}
\label{eq:logical vs physical xeb}
   \mb E_\beta \left[ f_\lint(Q_{\beta}, P_{\beta}) \right] &= 2^N \mb E_\beta \sum_x Q_{\beta}(x)P_{\beta}(x) -1, \\
    &= 2^N \mb E_\beta \sum_{x_b}\sum_{x_f} Q_{\beta}(x_f,x_b)P_{\beta}(x_f,x_b) -1, \nonumber \\
    &=2^n\mb E_\beta \sum_{x_b}Q_{\beta}(x_b) \sum_{x_f} Q_{\beta}(x_f|x_b)P_\beta(x_f|x_b) -1, \nonumber\\
    &=2^n \mb E_\beta \mb E_{x_b} \sum_{x_f} \tilde{Q}_{\beta, x_b}(x_f)  \tilde{P}_{\beta, x_b}(x_f) -1, \nonumber\\
    &= \mb E_\beta \mb E_{x_b}  \left[ \tilde f_\lint(\tilde{Q}_{\beta, x_b}, \tilde{P}_{\beta, x_b}) \right],\nonumber
\end{align} 
which we can rewrite in short notation as
\begin{equation}
    \overline f_\lint = \overline{\tilde{f}}_\lint
    \label{eq:equivalence logical physical linear xeb}.
\end{equation}
An analogous computation yields
\begin{equation}
\label{eq:shift log xeb}
     \overline{f}_\logt = \overline{\tilde{f}}_\logt + \log 2^{n(m-1)}.
\end{equation}

To summarize, the physical and logical average XEB fidelities are equivalent in both the linear and the logarithmic versions up to a shift for the log XEB fidelity. 
However, their empirical variance---given in \cref{eq:xeb total variance}---might still differ, because the samples are grouped differently in the mean of means estimator. 
In practice, we will therefore use XEB estimates from the physical circuits whenever possible given system size constraints (which influences the complexity of computing the ideal probabilities).
This is because we are able to take more samples per circuit (reducing the first term of \cref{eq:xeb total variance}) since the circuits of the logical XEB fidelity are partially determined by the---random---physical measurement outcomes. 
Moreover, in the experimental setting, taking more samples per circuit is cheaper than running more different circuits.

\paragraph*{Ideal values of the XEB fidelity.} 
For completeness, we conclude this subsection by demonstrating how to compute \textit{ideal values} of the average XEB fidelities.
These are the values of $\overline f_\lint $ and $ \overline f_\logt$ resulting from the case where $Q_\beta = P_\beta$, i.e.
\begin{align}
    \overline f_\lint(\mc P, \mc P ) &= \mb E_\beta \left[ f_\lint(P_\beta, P_\beta) \right] \quad \text { and }  \quad \overline f_\logt(\mc P, \mc P ) = \mb E_\beta \left[ f_\logt(P_\beta, P_\beta) \right].
\end{align}
To compute these ideal values we make use of statistical properties of the logical output distributions. In particular, it was shown by \textcite{haferkamp_closing_2020} that the logical circuits form an $\epsilon$-approximate $2$-design in depth $m \in O(n+ \log(1/\epsilon))$. 
This implies that the second moments $\mb E_{\beta, x_b} [ \tilde{P}_{\beta, x_b}^2 ]$ of the ideal logical output probability distributions approximate the Haar-random value with relative error $\epsilon$. 
Neglecting this error, we find that the ideal average linear XEB fidelity of circuits with such scaling of $m$ with $n$ asymptotically behaves as
\begin{align}
   \overline f_\lint(\mc P, \mc P ) 
   &= \overline{\tilde{f}}_\lint(\mc P, \mc P ), \nonumber\\
  & = 2^n  \sum_{x_f} \mb E_{\beta, x_b} \left[\tilde{P}_{\beta, x_b}(x_f)^2 \right] -1,\nonumber\\
  & =  \frac{2\cdot2^{2n}}{2^n(2^n+1)} -1 ,
  \nonumber\\
  & =  \frac 2{1 + 2^{-n}} -1 \approx 1 - \frac 1{2^{n-1}}. \label{eq:ideal value linear xeb}
\end{align}
In particular, because of the equality of physical and logical linear XEB, the ideal value can only depend on the size of the logical circuit which is in turn given by the shortest side of the square lattice.
Moreover, it is easy to see that the average linear XEB fidelity with the uniform distribution $\mathcal{U}$ (defined by $\beta \mapsto U([2^n]) \coloneqq (2^{-n}, 2^{-n}, \ldots, 2^{-n}) \in [0,1]^{2^n}$) is given by $\overline f_\lint(\mc U, \mc P ) =0$. 

We can repeat the same calculation for the logarithmic XEB fidelity,  to find 
\begin{align}
   \overline{f}_\logt(\mc P, \mc P )  
   &= \overline{\tilde{f}} _\logt(\mc P, \mc P ) + \log 2^{n(m-1)}, \nonumber\\
   & = - \sum_{x_f}  \mb E_{\beta,x_b} \left[ \tilde{P}_{\beta, x_b}(x_f) \log \tilde{P}_{\beta, x_b}(x_f) \right]  + \log 2^{n(m-1)},  \label{eq:ideal log -3} \\
   & =  \log 2^n - 1 + \gamma  + \log 2^{n(m-1)} ,  \label{eq:ideal log -2} \\
   & = \log 2^{nm} - 1 + \gamma\nonumber
\end{align}
where the step from \cref{eq:ideal log -3} to \eqref{eq:ideal log -2} follows from the properties of the exponential distribution (see Ref.~\cite[Sec.~II of the SI]{boixo_characterizing_2018} for details) and $\gamma \approx 0.5774 $ is the Euler constant. 
In particular, this calculation implies that for distributions with uniformly distributed marginals and Porter-Thomas distributed conditional probabilities, we get the same average value as for global Porter-Thomas distributed distributions.
Likewise, we find the average value of the log XEB fidelity when comparing it to the uniform distribution $\mathcal{U}$ to be 
\begin{align}
\label{eq:ideal value log uniform xeb}
\overline f_\logt(\mathcal{U}, \mathcal{P}) & = -  \Eb_\beta [\log P_\beta(x) ]  = \log 2^{nm} + \gamma, 
\end{align}
again, assuming Porter-Thomas shape of the distribution.

\begin{figure*}[t]
  \includegraphics{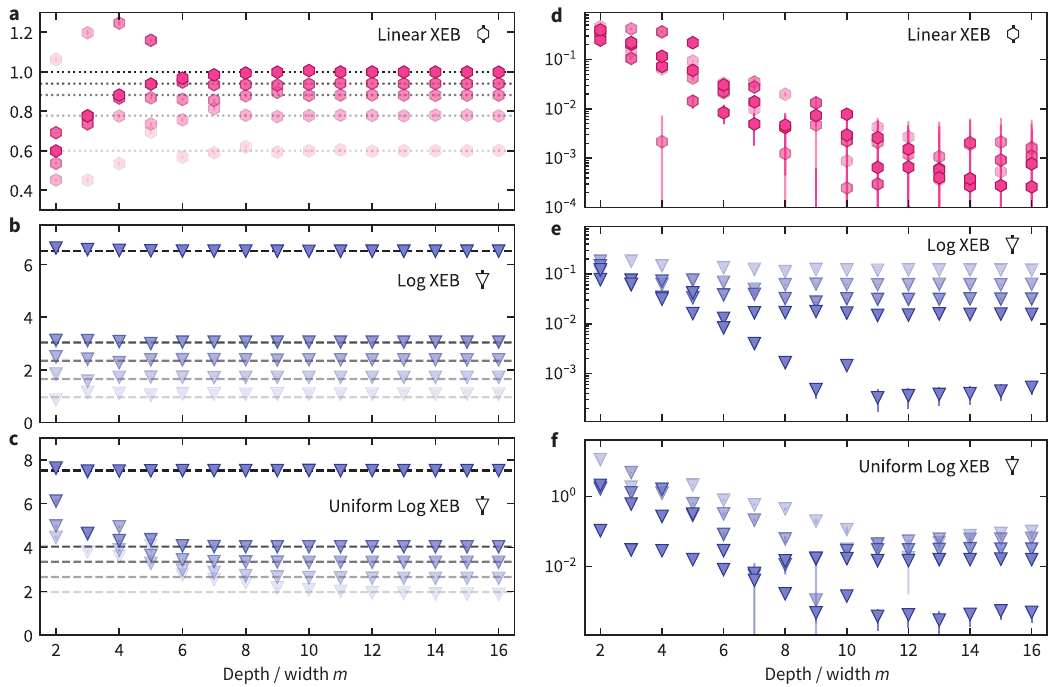}
  \caption{
  \label{fig:ideal values of the XEB}
  \textbf{Ideal values of the linear and logarithmic XEB fidelities.}
  We compute the ideal values of the XEB fidelities for clusters of size $n \times m$ with fixed values of the height $n = 2,3,4,5,10$ (corresponding to the number of qubits of the logical circuit) in increasing opacity and scaling of the width $m$ (corresponding to the depth of the logical circuit). 
  To this end, we average the XEB fidelities of the ideal output distributions of $K=10^5$ random (logical) circuits per data point. 
  Error bars are $3\sigma$ intervals.
  (a) Ideal values of the linear XEB (pink hexagons). For the linear XEB fidelity, the ideal value of the physical circuit equals that of the logical circuit by the uniformity property of the marginals, cf.\ \cref{eq:logical vs physical xeb}. 
   Dotted lines indicate the ideal values \eqref{eq:ideal value linear xeb}.
  (b) Ideal value of the logical log XEB fidelity $\overline{\tilde f}_\logt(\mc P, \mc P ) $ (blue triangles).
  Dashed lines represent the asymptotic value $\log 2^n + \gamma -1$ (\cref{eq:ideal log -2} \& \eqref{eq:shift log xeb}).
  (c) Logical log XEB fidelity for uniform samples $\overline{\tilde f}_\logt(\mc U, \mc P ) $. 
  Dashed lines indicate the ideal value $\log 2^n + \gamma$ (\cref{eq:ideal value log uniform xeb}).
  (d--f) Deviations of the estimated finite-size values from the ideal values of (a--c), respectively. 
  }
\end{figure*}

The theoretical ideal values found above pertain to the asymptotic limit of cluster states of increasing size. 
However, in the experiments reported in the main text, we deal with small instance sizes. 
In \cref{fig:ideal values of the XEB}, we confirm the convergence to the ideal values computed above for small instance sizes as relevant to our experiment.
We find that the linear XEB fidelity significantly deviates from the expected asymptotic value for small sizes $n \times m $ of the cluster. 
It is also true that the XEB fidelity of $n \times m $ clusters equals that of $m \times n $ clusters and is given by the ideal value of the logical circuit corresponding to the shorter side. 
When we measure the deviation of the ideal XEB fidelity of the logical circuits from their expected value, this deviation decays exponentially with the depth $m$ of the logical circuit (corresponding to the width of the cluster), independently of the number of qubits $n$ (corresponding to its height)---until we hit the noise floor set by the precision of our computation around $m = 10$.\footnote{
Note that in this comparison, we sometimes compare the XEB fidelities of rectangular clusters to the ideal XEB fidelity corresponding to the \emph{longer side}, namely whenever $n \geq m $.  } 

In contrast, the ideal logarithmic logical XEB is almost immediately close to its ideal value \eqref{eq:logical vs physical xeb}, but the deviation (after an initial decay) stays roughly constant with the width of the cluster while it decays with the height. 
We interpret this fact in terms of the Porter-Thomas distribution: 
For a small number of qubits $n$ on which the logical circuit acts, the exponential distribution is not a good approximation of the actual distribution of output probabilities of Haar-random quantum states.
Hence, the calculation of the mean value, which uses the exponential distribution, incurs a systematic error.\footnote{See the Supplementary Material of Ref.~\cite{hangleiter_anticoncentration_2018} for details of this calculation. }
Finally, the deviation of the ideal log XEB fidelity for uniform samples from the expected values decay with both the height and the width of the cluster. 
For the small-size experiments (up to $m=n=4$) we therefore need to use the computed values of the XEB fidelities (instead of the asymptotic values) when using XEB to estimate the fidelity; see the subsequent section.

\subsection{Estimating the average fidelity via XEB}
\label{ssec:average_fid_from_xeb}

In this section, we discuss how the average XEB fidelities $\overline f_\lint $ and $\overline f_\logt$ and their logical counterparts  $\overline{\tilde{f}}_\lint $ and $\overline{\tilde{f}}_\logt$ can be used to estimate the average state fidelity $\overline F$ of the underlying quantum states under assumptions on the noise in the device.
We follow the argumentation of \textcite[][Sec.~IV.A, SI]{arute_quantum_2019_short}, see also Ref.~\cite[][Sec.~V.B.3]{hangleiter_computational_2023} for an overview.

\subsubsection{Linear XEB fidelity with depolarizing noise}

Let us begin with the linear XEB fidelity. We consider first the toy model of global depolarizing noise and then generalize it to uncorrelated and unbiased noise.

\paragraph*{Depolarizing noise.}
Consider the noisy state 
\begin{align}
  \rho_\beta(\epsilon) = \epsilon \proj{\psi_\beta} + (1-\epsilon) \id/2^N. 
  \label{eq:noisy state depolarizing}
\end{align}
where $\ket{\psi_\beta}$ is the generalized cluster state.
Then, the fidelity is given by $F=\bra {\psi_\beta} \rho_\beta(\epsilon) \ket {\psi_\beta} = \epsilon + (1-\epsilon)/2^N$, and the same holds true for the average fidelity $\overline{F}$. 
In this case, the XEB fidelity is given by
\begin{align}
  f_\lint(Q_\beta(\epsilon),P_\beta) = \epsilon f_\lint (P_\beta, P_\beta) + (1-\epsilon) f_\lint(U([2^N]),P_\beta). 
\end{align}
Averaging over $\beta$, we find 
\begin{align}
    \overline f_\lint(\mc Q_, \mc P ) & =\Eb_\beta [f_\lint(Q_\beta(\epsilon),P_\beta)] \\
   & = \epsilon \Eb_\beta [f_\lint (P_\beta, P_\beta)] + (1-\epsilon) \underbrace{ \overline{f}_\lint(\mc U , \mc P)}_{=0}\nonumber\\ 
   & = \epsilon  \overline f_\lint(\mc P, \mc P ),
   \nonumber
\end{align}
and hence, we can estimate $\epsilon$ as
\begin{align}
\label{eq:epsilon estimate lin}
  \hat \epsilon = \frac{ \hat {\overline f}_\lint(\mc Q, \mc P )}{\overline f_\lint(\mc P, \mc P )}, 
\end{align}
where $\hat {\overline f}_\lint(\mc Q, \mc P )$ is the empirical estimate of the experimental average linear XEB fidelity.
We can then estimate the average fidelity as 
\begin{align}
\label{eq:physical fidelity estimator}
  \overline F(\hat \epsilon) = \hat \epsilon + (1- \hat \epsilon)/2^{nm}. 
\end{align}
\textcite{arute_quantum_2019_short} justify this estimator further using Bayes rule.

\paragraph*{Uncorrelated and unbiased noise.} 
One can make the same argument in case the quantum state is given by some noisy state
\begin{align}
  \rho_\beta(\epsilon) = \epsilon \proj{\psi_\beta} + (1- \epsilon) \chi_\beta,
  \label{eq:noisy state uncorrelated}
\end{align}
decomposed into the ideal state and a state $\chi_\beta$ capturing the noise. 
Now, the same conclusions regarding estimates of the \emph{average fidelity} will hold 
in case the noise is 
\begin{itemize}
  \item \emph{uncorrelated} in the sense that $\Eb_\beta[\bra x \chi_\beta \ket x \braket x {\psi_\beta} \braket{\psi_\beta} x] = \Eb_\beta[ \bra x \chi_\beta \ket x ]\Eb_\beta[\braket x {\psi_\beta} \braket {\psi_\beta} x] $, and
  \item \emph{unbiased} in the sense that  $\Eb_\beta[ \bra x \chi_\beta \ket x ] = 1/2^N$.  
\end{itemize}
The estimator $\hat {\overline F}$ using \cref{eq:epsilon estimate lin} will then give a good estimate of the average fidelity. 

We note that one can similarly relate the average logical linear XEB $\overline{\tilde{f}}_\lint $ to the average fidelity of the ``logical'' output state, i.e., the state on the final column that arises from measuring the bulk qubits obtaining some outcome $x_b$ which we denote by $\ket{\psi_{\beta,x_b}}$. To see this, consider this noisy logical state $\rho_ {\beta,x_b}(\tilde\epsilon)$ and write analogously to \Cref{eq:noisy state depolarizing}
\begin{equation}
    \rho_{\beta,x_b}(\tilde \epsilon) = \tilde \epsilon \proj{\psi_{\beta,x_b}} + (1-\tilde \epsilon) \id/2^n.
\end{equation}
Then, the average fidelity of the noisy logical output state is given by 
\begin{align}
\label{eq:logical fidelity estimator}  
\overline{\tilde F}(\tilde \epsilon) =  \tilde\epsilon +(1-\tilde\epsilon)/2^n.
\end{align}
In analogy to \Cref{eq:epsilon estimate lin}, this average logical fidelity can be estimated via the average logical XEB as follows 
\begin{align}
\label{eq:epsilon estimate logical lin}
  \hat {\tilde\epsilon}_\lint = \frac{ \hat {\overline{\tilde{f}}}_\lint(\mc Q, \mc P )}{\overline{\tilde{f}}_\lint(\mc P, \mc P )} = \hat {\epsilon}_\lint
\end{align}
where the last equality follows from the equivalence of physical and logical average linear XEB derived in \Cref{eq:equivalence logical physical linear xeb}.

\subsubsection{Log-XEB}

We can repeat the same argument for  the logarithmic XEB under the global depolarizing noise assumption, and find
\begin{align}
  \overline {f}_\logt(\mc Q_, \mc P ) &=\Eb_\beta [f_\logt(Q_\beta(\epsilon),P_\beta)] \\
 & = \epsilon \overline f_\logt(\mc P, \mc P )  +(1-\epsilon) \overline {f}_\logt(\mc U, \mc P)
\end{align}
and hence an estimator of $\epsilon$ 
is given by 
\begin{align}
\label{eq:epsilon estimate log}
  \hat \epsilon_\logt &= \frac{\hat { \overline {f}}_\logt(\mc Q, \mc P) - \overline {f}_\logt(\mc U, \mc P)  }{\overline {f}_\logt(\mc P, \mc P) -  \overline {f}_\logt(\mc U, \mc P)} .
\end{align}
Again, from $\hat{\epsilon}$, we can estimate the average (physical) fidelity according to $\overline F(\hat \epsilon)$. Again, we find $\hat \epsilon_\logt = \hat{\tilde \epsilon}_\logt$, since all logical logarithmic XEB quantities are just shifted by $\log 2^{n(m-1)}$. 
However, for the same derivation to work with more general noise, we need to adapt the ``uncorrelated'' assumption to the logarithm, i.e., 
$\Eb_\beta[\bra x \chi_\beta \ket x \log(P_\beta( x)] = \Eb_\beta[ \bra x \chi_\beta \ket x ]\Eb_\beta[ \log(P_\beta( x)] $, while the unbiasedness condition remains the same. 
Notice that all estimators above are unbiased since they are just linear in the empirical estimates of the XEB fidelities. 

\subsubsection{How to estimate fidelity}
\label{appsec:fidelity estimation}

While the estimates $\hat{\tilde \epsilon}$ and $\hat \epsilon$ always agree, the average fidelity estimators thus differ in the normalization of the correction term to the state fidelity in Eqs.~\eqref{eq:physical fidelity estimator} and \eqref{eq:logical fidelity estimator}. 
Which correction to the average depolarizing fidelity will yield a better estimate of the fidelity depends on how accurate the uncorrelated and unbiased noise assumption is for the logical versus the physical output state, or in other words, how well the model of Eqs.~\eqref{eq:noisy state depolarizing} and \eqref{eq:noisy state uncorrelated} applies to the corresponding states. 

\textcite{dalzell_random_2021} show that for local random circuits of at least logarithmic depth, local depolarizing noise approximately transforms into global depolarizing (white) noise at the level of the output state and thus build confidence in the validity of these assumptions.
Specifically, \textcite{dalzell_random_2021} prove that the white-noise assumption is approximately true in random circuits provided the physical noise is local and unbiased. 
In that case, the effective noise at the end of the circuit will be approximately depolarizing with an error scaling inversely with the number of gates. More precisely, they show that the normalized linear XEB $\overline f = \overline f_\lint(\mc Q, \mc P)/\overline f_\lint(\mc P , \mc P)$ between the noisy distribution $\mc Q$ and the ideal distribution $\mc P$ behaves as 
\begin{align}
  \label{eq:dalzell formula}
 \overline f(\eta) =  \ee^{-2S \eta} ,
 \end{align} 
where $S$ is the number of  two-qubit gates and $\eta $ is the probability of a local Pauli error on each qubit after a two-qubit gate. Moreover, for incoherent \emph{unital} noise, the noisy distribution approaches the uniform distribution at the same rate with an error given by $O( \overline f_\lint  \eta \sqrt S)$ all in the regime of $S \in \Omega( n \log n)$. 

Given that the statistical properties of random logical MBQC circuits behave completely analogously to those of random circuits in the circuit model, we would expect an analogous result to hold for the fidelity of the logical output state in MBQC. 
More precisely, random logical MBQC circuits behave like random universal circuits on the level of lower moments in the sense that random logical MBQC circuits also generate unitary $2$-designs  \cite{haferkamp_closing_2020} (and presumably polynomial designs as well). 
A possible caveat, however, is that in MBQC physical noise translates non-trivially into logical noise as considered for single-qubit circuits by \textcite{usher_noise_2017}. 
Thus, while we are unclear on the exact conditions on a local noise model, we do expect that an analogous result to that of \textcite{dalzell_random_2021} holds for logical MBQC circuits. 
In this case, the logical XEB fidelity will be a good measure of the quantum fidelity of the output of the logical circuit.
In this case, the fidelity will decay approximately according to \cref{eq:dalzell formula}.

Notice, though, that our direct estimate of the fidelity measures the fidelity of the \emph{physical output state} and hence we cannot experimentally certify that XEB yields quality estimates for the logical fidelity $\overline{\tilde F}$. 
But for the physical output state, we are much less confident in the validity of the uncorrelatedness and unbiasedness of the noise with the circuit. 
Indeed, a priori, there is no good reason to expect that the physical XEB fidelity estimator $\overline F(\hat \epsilon)$ matches the physical fidelity $\overline F$ accurately unless physical and logical average fidelity behave in the same way. 

We do find, however, that the estimator $\overline F(\hat \epsilon)$ works reasonably well as an estimate for the fidelity for local depolarizing noise, and also the noise we face in the experiment; see the following section. 
This suggests that the uncorrelated and unbiasedness assumption does in fact hold true for the physical circuit as well. 
Furthermore, it suggests, that as $n,m$ grow, the physical and logical physical fidelity converge. 
We leave a more detailed analysis of the effect of noise in MBQC on the estimates of fidelity to future work. 
In the following section, we will provide numerical evidence that, indeed, the XEB fidelities can be used to estimate the quantum fidelity in the presence of various types of local noise on the physical circuit. 

\subsection{A numerical study }
\label{ssec:xeb numerics}

\begin{figure*}[t]
 \includegraphics{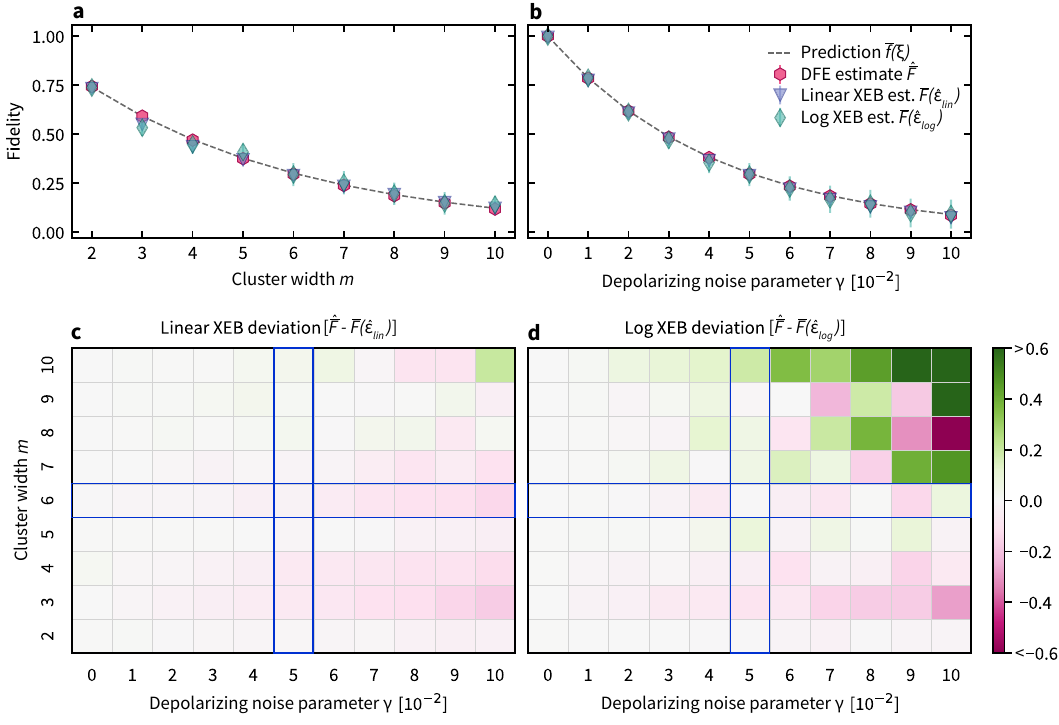}
  \caption{\textbf{Average fidelity estimation for depolarizing noise.} 
  We simulate sampling in the Hadamard basis and DFE for noisy random cluster states of size $2 \times m$ for $m \in \{2,3, \ldots, 10\}$ with local depolarizing noise with parameter $\gamma$ after every entangling $CZ$ gate. We sample $K = 10^5$ random circuits and take $M = 50$ samples per circuit. From the classical samples, we compute estimates of the average fidelity  $\overline F(\hat \epsilon)$ according to \eqref{eq:physical fidelity estimator} using the linear and logarithmic average XEB fidelity $\overline f_\lint$ and $\overline f_\logt$ (represented by purple triangles and green diamonds, respectively) and compare them to the DFE estimate $\hat {\overline F}$ of the average fidelity (pink hexagons) as well as the prediction \eqref{eq:dalzell formula} by \textcite{dalzell_random_2021} (dotted line).
  Error bars are $3\sigma$ intervals. 
  (a) Decay of the average fidelity  with the width of the cluster state for fixed noise parameter $\gamma = 0.05$.
  (b) Decay of the average fidelity with the noise parameter for a fixed width of the cluster $m = 6$ (blue boxes).
  (c) Deviation $\overline F(\hat \epsilon_\lint) - \hat {\overline F}$ between the fidelity estimate from the linear XEB fidelity and the average fidelity.  
  (d) Deviation $\overline F(\hat \epsilon_\logt) - \hat{ \overline F}$ between the fidelity estimate from the linear XEB fidelity and the average fidelity. 
  \label{fig:simulated_depolarizing}}  
  
\end{figure*}

\begin{figure*}[t]
 \includegraphics{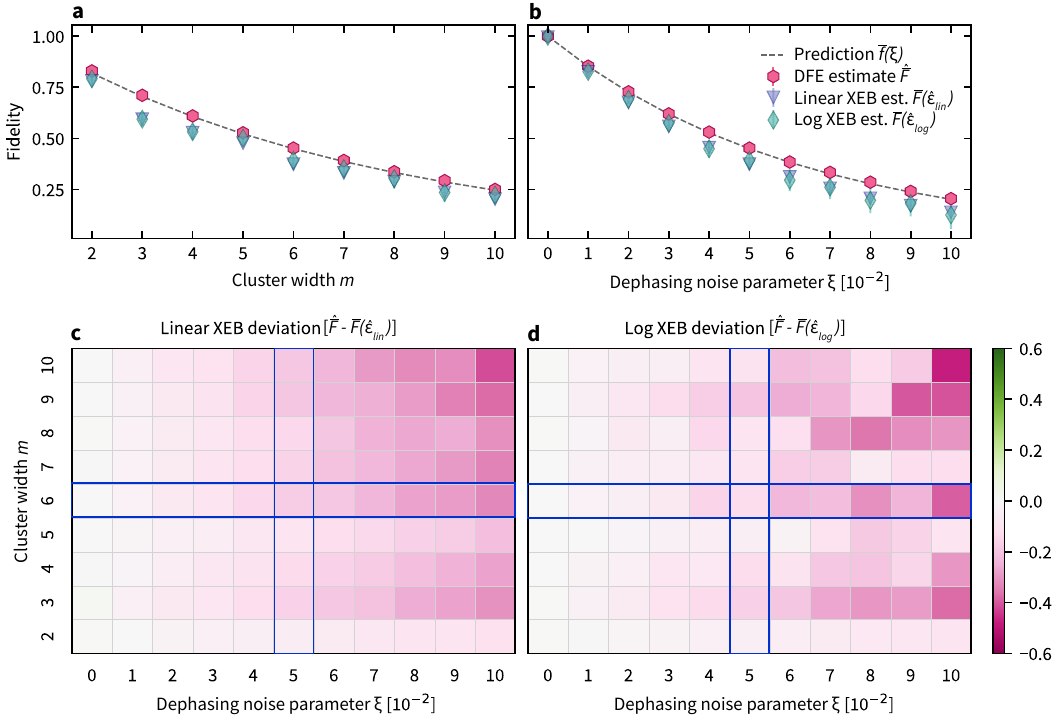}
  \caption{\textbf{Average fidelity estimation for dephasing noise.} 
  We simulate sampling in the Hadamard basis and DFE for noisy random cluster states of size $2 \times m$ for $m \in \{2,3, \ldots, 10\}$ with local dephasing noise with parameter $\xi$ after every entangling $CZ$ gate. We sample $K = 10^5$ random circuits and take $M = 50$ samples per circuit. From the classical samples, we compute estimates of the average fidelity  $\overline F(\hat \epsilon)$ according to \eqref{eq:physical fidelity estimator} using the linear and logarithmic average XEB fidelity $\overline f_\lint$ and $\overline f_\logt$ (represented by purple triangles and green diamonds, respectively) and compare them to the DFE estimate $\hat {\overline F}$ of the average fidelity (pink hexagons) as well as the prediction \eqref{eq:dalzell formula} by \textcite{dalzell_random_2021} (dotted line).
  Error bars are $3\sigma$ intervals. 
  (a) Decay of the average fidelity  with the width of the cluster state for fixed noise parameter $\xi = 0.05$.
  (b) Decay of the average fidelity with the noise parameter for a fixed width of the cluster $m = 6$ (blue boxes).
  (c) Deviation $\overline F(\hat \epsilon_\lint) - \hat {\overline F}$ between the fidelity estimate from the linear XEB fidelity and the average fidelity.  
  (d) Deviation $\overline F(\hat \epsilon_\logt) - \hat{ \overline F}$ between the fidelity estimate from the linear XEB fidelity and the average fidelity. .\label{fig:simulated_dephasing}}  
\end{figure*}

\begin{figure*}[t]
  \includegraphics{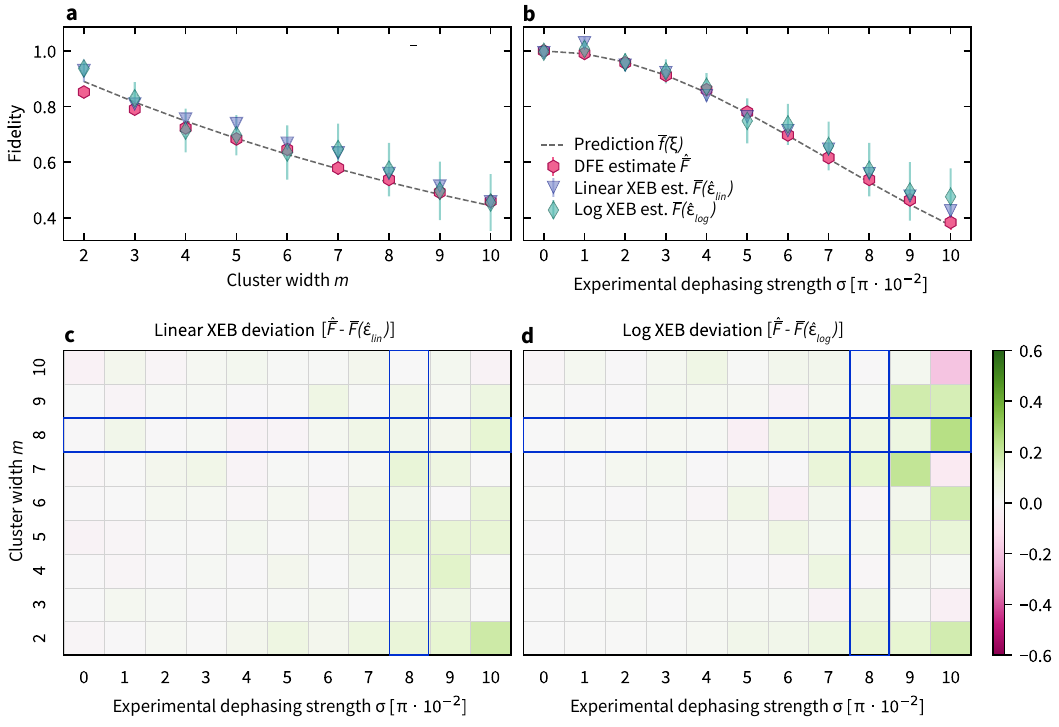}
  \caption{\textbf{Average fidelity estimation for the noisy experimental circuits.} 
  We simulate sampling in the Hadamard basis and DFE for noisy random cluster states of size $2 \times m$ for $m \in \{2,3, \ldots, 10\}$ with random local $Z$-rotations with rotation angle drawn from a normal distribution with mean $0$ and standard deviation $\sigma$ inserted in every layer of the experimental circuit; see the Methods section for details. We sample $K = 10^5$ random circuits and take $M = 1$ sample per circuit. 
  From the classical samples, we compute estimates of the average fidelity  $\overline F(\hat \epsilon)$ according to \eqref{eq:physical fidelity estimator} using the linear and logarithmic average XEB fidelity $\overline f_\lint$ and $\overline f_\logt$ (represented by purple triangles and green diamonds, respectively) and compare them to the DFE estimate $\hat {\overline F}$ of the average fidelity (pink hexagons) as well as the prediction $\overline f (3\gamma/4)$ \eqref{eq:dalzell formula} by \textcite{dalzell_random_2021} (dotted line).
  Error bars are $3\sigma$ intervals.
  (a) Decay of the average fidelity  with the width of the cluster state for fixed noise strength $\sigma = 0.08\pi$.
  (b) Decay of the average fidelity with the noise parameter for a fixed width of the cluster $m = 8$ (blue boxes).
  (c) Deviation $\overline F(\hat \epsilon_\lint) - \hat {\overline F}$ between the fidelity estimate from the linear XEB fidelity and the average fidelity.  
  (d) Deviation $\overline F(\hat \epsilon_\logt) - \hat{ \overline F}$ between the fidelity estimate from the linear XEB fidelity and the average fidelity. 
  \label{fig:simulated_experimental}}
\end{figure*}

In the previous sections, we have explained how to obtain estimates $\hat {\overline F} $ for the average fidelity of the state preparation of the cluster state in two different ways: 
\begin{itemize}
    \item From random stabilizer measurements on the cluster state via DFE as considered in \cref{ssec:average dfe}.
    \item From quantum random sampling in the Hadamard basis via XEB fidelity estimation as explained in  \cref{ssec:average_fid_from_xeb}.
\end{itemize}
In this section, we numerically study the quality of fidelity estimates obtained via XEB fidelity estimation by using the estimates obtained via DFE as a benchmark. 
In particular, we generate data according to the above-mentioned two methods by numerically simulating noisy state preparations $\rho_\beta$ of the cluster state for many randomly drawn $\beta$. 
We do so for different types of noise, system sizes, and noise strengths. 
From these data, we then obtain the corresponding average fidelity estimates as a function of the system size and noise strength.
More concretely, we numerically simulated two  different settings.

The first setting is inspired by the theoretical work of \textcite{dalzell_random_2021} on random circuit sampling under local, unbiased noise. 
This is the setting in which we most likely expect the average XEB fidelity estimate $\overline F(\hat \epsilon)$ to be a good estimate of the average fidelity $\overline F$.
Here, we consider noisy cluster state preparations $\rho_\beta$ via circuits built from Hadamard and $CZ$ gates and $Z$-rotation gates. 
We take all single-qubit gates to be perfect but all $CZ$ gates are followed by local depolarizing or dephasing noise channels, respectively. 
In \cref{fig:simulated_depolarizing} and \cref{fig:simulated_dephasing} we compare the fidelity estimates obtained via the DFE and XEB methods for these two noise settings. 
We also compare these results to fidelity scaling predicted via the formula \eqref{eq:dalzell formula}, where the number of two-qubit $CZ$ gates is  in our case just the number of edges of the $n \times m $ square lattice, given by $n (m-1)+ m(n-1)$. 
Writing the single-qubit depolarizing and dephasing channels $\mc D_\gamma$ and $D_\xi$ with parameters $\gamma$ and $\xi$, respectively, as 
\begin{align}
\label{eq:depolarizing channel}
  \mc D_\gamma(\rho) &  = (1- \gamma )\rho + \gamma \frac \id2 \\
  \label{eq:dephasing channel}
  D_\xi(\rho) &  = (1- \gamma )\rho + \xi \diag(\rho),
\end{align}
the error probability $\eta$ takes values $\eta = 3\gamma/4 $ for depolarizing noise and $\eta = \xi/2$ for dephasing noise.

We find excellent agreement of the prediction $\overline F(\eta)$ for the average physical fidelity---although it was derived for the XEB fidelity, while the XEB fidelity estimators are approximately correct for depolarizing noise in the regime of low noise parameters $\gamma \sim 1/n$. 
For dephasing noise, we find that the XEB fidelity estimators greatly underestimate the average fidelity. 

In contrast, the second setting models the actual experimental setup reported in the main text. That is, we simulate the noisy experimental circuits described in the Methods section. 
Again, we find excellent agreement of the physical fidelity with the prediction $\overline F(\eta)$ by \textcite{dalzell_random_2021}. We find that setting the effective depolarizing noise parameter $\gamma = 1 - \exp(- 0.310 \sigma^2/2)$, where $\sigma $ is the measure of the noise strength, gives the best fit with the observed fidelity.

\section{Error analysis for the mean of means estimator}
\label{app:mean error}

In order to compute the statistical error associated with our estimates of the fidelity and the XEB fidelity, we need to compute the variance of a finite-sample estimator of a random variable $A$ conditioned on a random variable $B$ so that
\begin{align}
  F = \mb E_{B \sim \mc B} \left [ \mb E_{A \sim \mc A} [A|B]\right].
\end{align}
We think of $B$ as the random circuit and $A$ as the samples or random stabilizer values of the random circuit.
Concretely, the fidelity estimate, average fidelity estimate \eqref{eq:average dfe},  and XEB fidelity \eqref{eq:linear xeb} estimate are obtained as the empirical estimate of the expectation values
\begin{align}
  F(\rho, \proj \psi) & = \frac 1 {2^N} \sum_{s \in \mc S} \sum_{\sigma \in \pm 1} \sigma \cdot \tr[\rho \pi_s^\sigma] \\
  & = \mb E_{s \sim \mc S}\left[ \mb E_{\sigma \sim p_{\rho,s}} [\sigma] \right] ,\nonumber \\
 \overline F &= \frac 1{2^{N}} \frac 1 {8^{N}} \sum_{\beta \in \frac\pi 4 \cdot [8]^n } \sum_{s_\beta \in \mc S_\beta} \sum_{\sigma = \pm 1} \sigma \cdot\langle\pi_{s_\beta}^\sigma \rangle_{\rho_\beta}\\  
  & = \mb E_{\beta, s_\beta \sim \mc S_\beta}\left[ \mb E_{\sigma \sim p_{\rho_\beta,s_\beta}} [\sigma] \right] , \\
  \overline f_\lint & = \mb E_{\beta \sim \frac \pi  4 \cdot [8]^N }\left[ \frac 1 {2^N} \sum_{x \in \{ 0,1 \}^N} Q_\beta(x) \left( 2^N P_\beta(x) - 1 \right) \right] \\
  & = \mb E_{\beta \sim \frac \pi 4 \cdot [8]^N } \left[\mb E_{x \sim Q_\beta} \left[2^N P_\beta(x) -1  \right] \right] , \nonumber
\end{align}
where $p_{\rho,s}(\sigma) = \tr[\rho \pi_s^\sigma]$ and $s = \pi_s^+ - \pi_s^-$ is a stabilizer of $\ket \psi$. 
We now wish to estimate the variance of a finite-sample estimate of such an expectation value, that is, an estimator 
\begin{align}
  \hat F = \frac 1 K \frac 1 M \sum_{i = 1}^K \sum_{j = 1}^M A_{i,j}, 
\end{align}
where $K$ is the number of times the first expectation value is sampled out, and $M$ is the number of times the second expectation value is sampled out, given the result of the first. 
For instance, to estimate the average bias of a bag of coins with different biases, we draw $K$ coins and flip each drawn coin $M$ times. In this case, $A_{i,j} \in \{0,1\}$ is the outcome of the $j$th flip of the $i$th drawn coin. 

Very generally, we can compute the variance of such a conditional expectation value using the law of total variance, which states that for two random variables $A,B$ on the same probability space 
\begin{align}
  \var(A) = \mb E[ \var [A| B ]] + \var[ \mb E [ A | B ]]. 
\end{align}

Consider the fidelity estimate.  
If we let $A_{i} = \sum_{j=1}^M A_{i,j}$ be the random variable, representing the empirical cumulative value of the $M$ measurement outcomes $A_{i,j} \in \{\pm1\}$ of  the stabilizer $s_i$ on the state preparation $\rho$ with associated probability distribution $p_{s_i,\rho}$, then the overall variance is given by $\var(\hat F) = (KM)^{-2}\sum_{i=1}^K \var(A_i)$. To understand the variance $\var(A_i)$, letting $p_i = p_{s_i, \rho}(+1)$ so that $\mb E[A_i] =  M (2 p_i -1) $, 
we now invoke the law of total variance, to get
\begin{align}
  \var[A_i]&  = \mb E_{p_i}\left[ \var [A_i |p_i]\right] + \var_{p_i} \left[\mb E[A_i |p_i] \right]  \\
  & = \mb E[ 4 M p_i (1 - p_i )]  + \var [M (2 p_i - 1) ]
  \nonumber \\
  & = 4 M \mb E[ p_i] - 4M \mb E[ p_i^2]  + 4 M^2 \var [p_i]
  \nonumber \\
   & = 4 M\mb E[ p_i] - 4 M(\var[p_i] + \mb E[ p_i]^2)   + 4M^2\var [p_i]\nonumber \\
   & = 4 M(\mb E[p_i] (1 - \mb E[p_i])) + 4M(M-1) \var[p_i].
   \nonumber 
\end{align}
This yields the overall variance
\begin{align}
  \var[\hat F] &  = \frac 4 {KM} \left( (\mb E[p_i] (1 - \mb E[p_i])) + (M-1) \var[p_i]\right)\\
  & = \frac 4 {KM}   (\mb E[p_i] (1 - \mb E[p_i])) + \frac 4 {K}\left( 1-\frac 1M \right) \var[p_i] . \nonumber
\end{align}
Consequently, the variance is asymptotically dominated by the variance over the stabilizers, but in the finite-sample case, there is a trade-off between choosing $M $ and $K$ governed by the specific value of $\var[p_i]$. 

For the case of the (linear and logarithmic) XEB fidelity estimated from $K$ random choices of rotation angles $\beta_i$, $i=1, \ldots, K$ and $M$ samples per $\beta_i$, we follow the same reasoning, defining the average estimator 
$\hat{ \overline f}_\lint =(KM)^{-1} \sum_{i,j=1}^{K,M} (2^N P_{\beta_i}(x_j) -1)$, and the single-circuit estimator $\hat{f}_\lint(Q_\beta, P_\beta) =M^{-1} \sum_{j=1}^{M} (2^N P_{\beta_i}(x_j) -1)$. 
We then consider the random variable $A_i = \sum_{j=1}^M(2^N P_{\beta_i}(x_j) -1)$ and $\var [\hat{ \overline f}_\lint] = (KM)^{-2} \sum_{i=1}^K \var[A_i]$,
and find
\begin{align}
  \var[A_i] = \mb E[ \var [A_i|\beta_i]] + \var [\mb E[ A_i|\beta_i]], 
\end{align}
and estimate the separate terms as 
\begin{align}
   \mb E_i[ \var_x [A_i|\beta_i]] & = \mb E_i[ M\var_x[ 2^N P_{\beta_i}(x)-1]]\nonumber\\
   & = \mb E_i[ \sum_x Q_{\beta_i}(x) (2^N P_{\beta_i}(x)-1 - f_\lint(Q_{\beta_i}, P_{\beta_i})))^2] \nonumber \\
   & \approx \frac 1 K \sum_{i=1}^K \sum_{j = 1}^M (2^N P_{\beta_i}(x_j)-1 - \hat f_\lint(Q_{\beta_i}, P_{\beta_i}))^2,
   \label{eq:estimate xeb expectation variance}\\
   \var_i [\mb E_x[ A_i|\beta_i]] & = \var_i[M f_\lint(Q_{\beta_i}, P_{\beta_i})]\nonumber\\
   & \approx M^2 \frac 1 K \sum_{i=1}^K(\hat f_\lint(Q_{\beta_i}, P_{\beta_i}) - 
   \hat{ \overline f}_\lint(Q_{\beta_i}, P_{\beta_i}))^2.
   \label{eq:estimate xeb variance expectation}
\end{align}
Overall, we obtain
\begin{align}
  \var[\hat {\overline f}_\lint ] & = \frac 1 {K^2} \frac 1 {M^2} 
    \left(\sum_{i=1}^K   \sum_{x=1}^M (2^N P_{\beta_i}(x)-1 - f_\lint(Q_{\beta_i}, P_{\beta_i}))^2 
  + M^2  \sum_{i=1}^K (f_\lint(Q_{\beta_i}, P_{\beta_i}) -  \overline f_\lint(Q_{\beta_i}, P_{\beta_i}))^2 \right) \nonumber \\
  &=\frac 1 {KM} \mb E_\beta [ \var_x[2^N P_{\beta}(x)-1]] + \frac 1 K \var_\beta [f_\lint(Q_{\beta}, P_{\beta})],\label{eq:xeb total variance}
\end{align}
which we estimate using the expressions in Eqs.~\eqref{eq:estimate xeb expectation variance} and \eqref{eq:estimate xeb variance expectation}. 
An analogous expression gives the variance of the logarithmic XEB estimate. 
\putbib
\end{bibunit}

\end{document}